\documentclass[10pt]{article}
\usepackage{amsfonts,color}
\usepackage{amsmath}
 \usepackage{epsfig}
\usepackage{lscape}
\usepackage{amssymb}
\usepackage{graphicx}
\usepackage{makeidx}
\usepackage{amssymb}
 \usepackage{footnote}

\DeclareMathOperator{\curl}{curl}
\DeclareMathOperator{\Div}{Div}

\DeclareMathOperator{\Curl}{Curl}

\newcommand{\Co}{C_0^{\infty}(\Omega)}

\newcommand{\norm}[1]{\|#1\|}

\newcommand{\R}{\mathbb{R}}

\newcommand{\C}{\mathbb{C}}

\renewcommand{\skew}{\mathop{\rm skew}}

\DeclareMathOperator{\sym}{sym}

\DeclareMathOperator{\dev}{dev}
\DeclareMathOperator{\sL}{\mathfrak{sl}}
\DeclareMathOperator{\so}{\mathfrak{so}}
\DeclareMathOperator{\gl}{\mathfrak{gl}}

\newcommand{\Sym}{ {\rm{Sym}} }

\newcommand{\Mprod}[2]{ {\langle #1 ,#2\rangle} }
\newcommand{\id}{ {1\!\!\!\:1 } }

\newcommand{\tr}[1]{ {\Tr \left[{#1}\right]} }

\allowdisplaybreaks[1]

\makeindex

\newtheorem{theorem}{Theorem}[section]

\newtheorem{corollary}[theorem]{Corollary}

\newtheorem{lemma}[theorem]{Lemma}

\newtheorem{proposition}[theorem]{Proposition}

\def\barr{\begin{array}}
\def\id{1\!\!1}
\def\tr{\textrm{tr}}

\def\dvg{\textrm{Div}}
\def\crl{\textrm{Curl}}

\def\dd{\displaystyle}

\def\barr{\begin{array}}
\def\earr{\end{array}}
\def\bec#1{\begin{equation}\label{#1}}
\def\becn{\begin{equation*}}
\def\endec{\end{equation}}
\def\endecn{\end{equation*}}
 \def\C{\mathbb{C}}
 \def\H{\mathbb{H}}
  \def\L{\mathbb{L}_c}
  \def\cdp{\,\!\cdot\!\,}
\def \e{{\mathrm{e}}}
\def \pd{\partial}

\newtheorem{remark}{Remark}[section]
\begin{document}

\title{The relaxed linear  micromorphic continuum:
well-posedness of the static problem and relations to the gauge theory of dislocations}
\author{
Patrizio Neff\thanks{Patrizio Neff,  \ \ Head of  Lehrstuhl f\"{u}r Nichtlineare Analysis und Modellierung, Fakult\"{a}t f\"{u}r Mathematik, Universit\"{a}t Duisburg-Essen,  Thea-Leymann Str. 9, 45127 Essen, Germany, email: patrizio.neff@uni-due.de} \quad
and \quad
Ionel-Dumitrel Ghiba\footnote{Ionel-Dumitrel Ghiba, \ \ Lehrstuhl f\"{u}r Nichtlineare Analysis und Modellierung, Fakult\"{a}t f\"{u}r Mathematik, Universit\"{a}t Duisburg-Essen, Thea-Leymann Str. 9, 45127 Essen, Germany;   Alexandru Ioan Cuza University of Ia\c si, Department of Mathematics,  Blvd. Carol I, no. 11, 700506 Ia\c si,
Romania;  Octav Mayer Institute of Mathematics of the
Romanian Academy, Ia\c si Branch,  700505 Ia\c si; and   Institute of Solid Mechanics, Romanian Academy, 010141 Bucharest, Romania, email: dumitrel.ghiba@uni-due.de, dumitrel.ghiba@uaic.ro}\quad
and \quad
Markus Lazar \footnote{Markus Lazar,\ \ Heisenberg Research Group, Department of Physics, Darmstadt University of Technology, Hochschulstr. 6, D-64289 Darmstadt, Germany, email:  lazar@fkp.tu-darmstadt.de}
  \quad
and \quad\\
Angela Madeo\footnote{Angela Madeo:  \ \  Laboratoire de G\'{e}nie Civil et Ing\'{e}nierie Environnementale, Universit\'{e} de Lyon-INSA, B\^{a}timent Coulomb, 69621 Villeurbanne Cedex, France;\ \ International Center M\&MOCS ``Mathematics and Mechanics of Complex Systems", Palazzo Caetani,
Cisterna di Latina, Italy, email: angela.madeo@insa-lyon.fr} }
\maketitle

\begin{abstract}
In this paper we consider the equilibrium problem in the relaxed linear model of  micromorphic elastic materials. The basic kinematical fields of this extended continuum model are the displacement $u\in \mathbb{R}^3$ and the non-symmetric  micro-distortion density tensor $P\in \mathbb{R}^{3\times 3}$. In this relaxed theory a symmetric force-stress tensor arises despite the presence of microstructure and the curvature contribution depends solely  on the micro-dislocation tensor $\Curl P$. However, the relaxed model is  able to fully describe rotations of the microstructure and to predict non-polar size-effects.  In contrast to classical linear micromorphic models, we allow the usual elasticity tensors  to become positive-semidefinite.  We prove that, nevertheless, the equilibrium problem has a unique weak solution in a suitable Hilbert space. The mathematical framework also settles the question of which boundary conditions to take for the micro-distortion. Similarities and differences between
linear micromorphic elasticity and dislocation gauge theory are discussed and
pointed out.
\\
\vspace*{0.25cm}
\end{abstract}

\newpage

\tableofcontents

\section{Introduction}

In this paper we consider  {\it the static variant of the relaxed micromorphic model} introduced in \cite{NeffGhibaMicroModel}. This new model reconciles  Kr\"{o}ner's rejection \cite{Kroner} of antisymmetric force stresses in dislocation theory with the asymmetric dislocation model of Eringen and Claus \cite{Eringen_Claus71} and  shows that the concept of asymmetric force stress is not strictly needed in order to describe rotations of the microstructure in non-polar materials. In fact, a non symmetric local force stress tensor $\sigma$ deviates considerably from classical linear elasticity theory and indeed it does not necessarily appear in gradient elasticity \cite{NeffGhibaMicroModel}. After more than half a century of intensive research there is no conclusive
experimental evidence for the necessity of non-symmetric force stresses,
at least for what concerns a huge variety of natural and engineering
micro-structured materials. However, some very special engineering
meta-materials like phononic crystals and lattice structures may need
the introduction of asymmetric force stress to fully describe their
mechanical behavior. This fact was observed in \cite{MadeoNeffGhibaW}, in
which the presence of the Cosserat couple  modulus $\mu_{c}>0$ has been
proved to be necessary for the physically correct description of the dynamical
behavior of high-tech micro-structured materials which are known to
show frequency band-gaps in the dynamic regime. The relaxed micromorphic
model with positive Cosserat modulus proposed in \cite{MadeoNeffGhibaW} is the only
generalized continuum model which is able to predict frequency band-gaps
contrarily to what is possible in the Mindlin-Eringen model (see \cite{Mindlin63,Mindlin64,Eringen64,Eringen99})
or in so-called second-gradient models (see e.g. \cite{2ndGradWaves,2nDGradSurface,Guyader,PhilippeBoundaryLayer}).
The described  new meta-materials are able to stop wave propagation when excited
with signals at frequencies falling in a precise range of values.
Materials of this type are expected to have important technological
applications for what concerns control of vibrations and would provide
a valid alternative to currently used piezo-electric materials which
are vastly studied in the literature (see e.g. \cite{Piezo1,isola2,isola6,isola1,isola3,Piezo}). Therefore, it gets clear that the asymmetry of the force stress tensor in a continuum theory is not an immediate consequence  of the presence of microstructure in the body, it is rather a constitutive assumption \cite{Bors}.
Thus, in the relaxed model we dispose of this assumption. Despite the simplification of giving rise to symmetric stress, the relaxed model preserves full kinematical freedom (\textit{12 degree of freedom}). Moreover, the proposed relaxed model is still able to fully describe rotations of the microstructure and to fit a huge class of mechanical behaviours of materials with microstructure. Another strong point of the relaxed theory is  that for the isotropic relaxed micromorphic model only \textit{3 curvature parameters} remain to be determined, which may eventually be reduced to 2 parameters, which are needed for fitting bending and torsion experiments (see e.g. \cite{Neff_STAMM04,Neff_Forest_jel05}). This could be a decisive step in the main problematic of determination of constitutive parameters in the micromorphic theory  of elastic materials (and other more general extended continuum models). The relaxed formulation of micromorphic elasticity has some similarities to recently studied models of gradient plasticity \cite{ForestJMPS10,Forest,Forest06,Forest02b,ForestTrinh11}.

The mathematical analysis of general micromorphic solids is well-established for infinitesimal, linear elastic models, see, for example \cite{Soos69,Hlavacek69,IesanNappa2001,Iesan2002}. The only known
existence results for the static geometrically nonlinear formulation are due to Neff
\cite{Neff_micromorphic_rse_05} and to Mariano and Modica \cite{Mariano08a}. Compared with \cite{Neff_micromorphic_rse_05}, Mariano and Modica \cite{Mariano08a} assume much more stringent coercivity assumptions which restrict the material response. As for the numerical implementation, see  e.g. \cite{Mariano05} and the development in \cite{Klawonn_Neff_Rheinbach_Vanis09}. In \cite{Klawonn_Neff_Rheinbach_Vanis09}
the original problem  is decoupled into two separate problems and the corresponding domain-decomposition techniques for the subproblem related to balance of forces are investigated. On the other hand, in the  classical linear theory of Mindlin-Eringen micromorphic elasticity, existence and uniqueness results were already established by S\'{o}os \cite{Soos69}, by Hlav{\'a}{\v c}ek \cite{Hlavacek69},  by Ie\c san and Nappa \cite{IesanNappa2001} and by Ie\c san \cite{Iesan2002} assuming that the free energy is  a pointwise positive definite quadratic form in terms of the usual set of independent constitutive variables \cite{Mindlin64,Eringen64,Eringen99}. Ie\c san \cite{Iesan2002} also gave a uniqueness result for the dynamic problem without assuming that the free energy is  a positive definite quadratic form.
Moreover, in order to study the existence of solution of the resulting  system, Hlav{\'a}{\v c}ek \cite{Hlavacek69}, Ie\c san and Nappa \cite{IesanNappa2001} and Ie\c san \cite{Iesan2002} considered the  strong anchoring boundary condition: the micro-distortion $P$ is completely described at a part of the boundary.  In contrast with the models considered until now, our
free energy of the relaxed model is {\it not uniformly pointwise positive definite} in the
control of the classical constitutive variables. We mention that the well-posedness  of the dynamic problem of the relaxed model is established in \cite{NeffGhibaMicroModel}, while a well-posed problem class of autonomous evolutionary equations  from elasticity theory modeling solids with micro-structure are studied in \cite{picard}.

As far as the mechanical behavior of micro-structured materials in
the static regime is considered, no clear experimental evidence about
the real need of introducing a non-symmetric force stress has been
provided up to now. Nevertheless, some rare experiments exist on special
micro-structured materials which are subjected to particular loading
and boundary conditions which would show the need of Cosserat theory
for the complete description of their mechanical behavior also in
the static case. This is what is described e.g. in \cite{YangLakes} in
which experimental evidence of the need of Cosserat elasticity seems
to be provided for what concerns small samples of compact bone under
torsion. In \cite{YangLakes} it is shown that as soon as the size
of the specimen becomes so small that the influence of deformation
of bone-microstructure (osteons) on macroscopic overall deformation
cannot be neglected, then the use of Cosserat theory becomes necessary
for the correct description of deformation of such materials. This
need is related to the fact that the micro-deformation associated
to relative rotations of osteons inside the considered specimens becomes
comparable to the overall macroscopic deformation of the sample. It
is clear that, in order to activate the deformation modes which need
a Cosserat-type theory for their correct description, multiple conditions
must be simultaneously satisfied (specimens of small sizes, applied
external loads and boundary conditions which excite particular micro-rotations,
etc.). In all other cases a symmetric force stress tensor would be
sufficient to correctly describe the mechanical behavior of the same
material. We can hence conclude that, except for what happens in some
particular cases in which the relative micro/macro-rotations cannot
be neglected, the use of a relaxed model is fully sufficient for the
description of the mechanical behavior of materials with microstructure,
both in the static and the dynamic case. We explicitly remark that the proposed relaxed model is thought for
applications involving isotropic materials undergoing small deformations.

The main point of the present work is to prove that the static problem of the new micromorphic relaxed model  \cite{NeffGhibaMicroModel} is {\it well-posed}, i.e. we study the existence and uniqueness of the solution in absence of inertial effects, and moreover we point out the relation with the dislocation gauge theory.  All the results are obtained for
 a standard set of {\it tangential boundary conditions} for the micro-distortion $P$, i.e. $P.\,\tau=0$ ($P_i\times\,n=0$, $i=1,2,3$) on $\partial \Omega$ and not the {\it usual strong anchoring condition} $P=0$ on $\partial \Omega$.  The solution space for the elastic distortion and micro-distortion is only ${\rm H}_0(\Curl;\Omega)\supsetneq {\rm H}^1_0(\Omega)$ (see \cite{Raviart79}) and for the macroscopic displacement $u\in {\rm H}^1_0(\Omega)$.   As mentioned, compared to other existence results established in the static case of the theory of micromorphic elastic materials,  we allow the usual elasticity tensors  to become positive-semidefinite. The main point in establishing the desired estimates is represented by the {\it new coercive inequalities} recently proved by Neff, Pauly and  Witsch \cite{NeffPaulyWitsch,NPW2,NPW3} and by Bauer, Neff, Pauly and Starke \cite{BNPS2,BNPS3} (see also \cite{LNPzamp2013}).

The relaxed formulation of micromorphic elasticity has some similarities
to the {\it gauge theory of dislocations}
given by Lazar~\cite{Lazar00,LazarJPMG02,Lazar02},
Lazar and Anastassiadis~\cite{Lazar2008,Lazar2009}
and Agiasofitou and Lazar~\cite{AL10}.
In fact, in \cite{Lazar2009,Lazar-maugin}
a simplified static version of  the isotropic  Eringen-Claus  model for
dislocation dynamics \cite{Eringen_Claus69}
has been investigated with $\mathbb{H}=0$ and $\mu_c\ge0$,
with a focus on the gauge theory of dislocations.
However, the dynamical theory of Lazar \cite{Lazar2008,Lazar2009b,Lazar10,Lazar-MMS11}
cannot be derived from Mindlin's dynamic theory,
since  in \cite{Lazar2008} there appears a (dynamical) gauge potential
which has no counterpart in Mindlin's model.
In the dislocation gauge theoretical formulation
dislocations arise naturally as a consequence of broken translational
symmetry and therefore their existence is not required to be postulated a priori. In the last part of this paper we explain the similarities and the differences between the relaxed micromorphic elastic theory and the gauge theory of dislocations.

\section{Formulation of the problem}

\subsection{Notation and  main inequalities}
For $a,b\in\R^3$ we let $\Mprod{a}{b}_{\R^3}$  denote the scalar product on $\R^3$ with
associated vector norm $\norm{a}_{\R^3}^2=\Mprod{a}{a}_{\R^3}$.
We denote by $\R^{3\times 3}$ the set of real $3\times 3$ second order tensors which are denoted in the sequel by capital letters.
The standard Euclidean scalar product on $\R^{3\times 3}$ is given by
$\Mprod{X}{Y}_{\R^{3\times3}}=\tr({X Y^T})$, and thus the Frobenius tensor norm is
$\norm{X}^2=\Mprod{X}{X}_{\R^{3\times3}}$. In the following we omit the index
$\R^3,\R^{3\times3}$. The identity tensor on $\R^{3\times3}$ will be denoted by $\id$, so that
$\tr({X})=\Mprod{X}{\id}$. By  $T(3)$ we denote  the three-dimensional
translation group.
We let $\Sym(3)$  denote the set of symmetric tensors. We adopt the usual abbreviations of Lie-algebra theory, i.e.,
 $\so(3):=\{X\in\mathbb{R}^{3\times3}\;|X^T=-X\}$ is the Lie-algebra of  skew symmetric tensors
and $\sL(3):=\{X\in\mathbb{R}^{3\times3}\;| \tr({X})=0\}$ is the Lie-algebra of traceless tensors.
 For all $X\in\mathbb{R}^{3\times3}$ we set $\sym X=\frac{1}{2}(X^T+X)\in\Sym(3)$, $\skew X=\frac{1}{2}(X-X^T)\in \so(3)$ and the deviatoric part $\dev X=X-\frac{1}{3}\;\tr({X})\,\id\in \sL(3)$  and we have
the orthogonal Cartan-decomposition  of the Lie-algebra $\gl(3)$
\begin{align}\label{odesc}
\gl(3)&=\{\sL(3)\cap \Sym(3)\}\oplus\so(3) \oplus\mathbb{R}\!\cdot\! \id,\notag\\
X&=\dev \sym X+ \skew X+\frac{1}{3}\tr(X)\!\cdot\! \id\,.
\end{align}

We consider a micromorphic continuum which occupies a bounded domain $\Omega$ and is bounded by the piecewise smooth
surface $\partial \Omega$.  The equilibrium of the body is referred to a fixed system of rectangular Cartesian axes $Ox_i$, $(i=1,2,3)$. Throughout this paper (when we do not specify else) Latin subscripts take the values $1,2,3$.
The micro-distortion (plastic distortion) $P=(P_{ij}):\Omega\rightarrow \mathbb{R}^{3 \times 3}$  is intended to describe the substructure of the material which can rotate, stretch, shear and shrink, while  $u=(u_i) :\Omega\rightarrow  \mathbb{R}^3$  is the displacement of the macroscopic material points. Typical conventions for differential
operations are implied such as comma followed
by a subscript to denote the partial derivative with respect to
 the corresponding cartesian coordinate.

The quantities involved in our new relaxed micromorphic continuum  model have the following physical signification:
\begin{itemize}
\item $(u,P)$ are the {\it kinematical variables},
\item ${\sigma}$ is  the force-stress tensor (the  Cauchy stress tensor, second order, symmetric),
\item $s$ is the microstress tensor (second order, symmetric),
\item ${m}$ is the moment stress  tensor (micro-hyperstress tensor, third order,  in general non-symmetric),
\item $u$ is the displacement vector (translational degrees of freedom),
\item $P$ is the micro-distortion tensor (``plastic distortion", second order, non-symmetric),
\item $f$ is the external body force,
\item $M$ is the external body moment tensor (second order, non-symmetric),
\item $e:=\nabla u-P$ is the elastic distortion (relative distortion or gauge potential, second order, non-symmetric),
\item $\varepsilon_e:=\sym e=\sym(\nabla u-P)$ is the elastic strain tensor (second order, symmetric),
\item $\varepsilon_p:=\sym P $ is the micro-strain tensor (``plastic strain", second order, symmetric),
\item  $\alpha:=\Curl e=-\Curl P$ is the micro-dislocation density tensor (translational field strength, second order).
\end{itemize}

By $\Co$ we denote  the set of smooth functions with compact support in $\Omega$.
 The usual Lebesgue spaces of square integrable functions, vector or tensor fields on $\Omega$ with values in $\mathbb{R}$, $\mathbb{R}^3$ or $\mathbb{R}^{3\times 3}$, respectively will be denoted by $L^2(\Omega)$. Moreover, we introduce the standard Sobolev spaces \cite{Adams75,Leis86,Raviart79}
\begin{align}
&{\rm H}^1(\Omega)=\{u\in L^2(\Omega)\, |\, {\rm grad}\, u\in L^2(\Omega)\}, \ \ \ {\rm grad}=\nabla\, ,\notag\\\notag
&\ \ \ \ \ \ \ \ \ \ \ \ \ \ \ \|u\|^2_{{\rm H}^1(\Omega)}:=\|u\|^2_{L^2(\Omega)}+\|{\rm grad}\, u\|^2_{L^2(\Omega)}\, ,\\
&{\rm H}({\rm curl};\Omega)=\{v\in L^2(\Omega)\, |\, {\rm curl}\, v\in L^2(\Omega)\}, \ \ \ {\rm curl}=\nabla\times\, ,\\\notag
&\ \ \ \ \ \ \ \ \ \ \ \ \ \ \ \ \ \ \|v\|^2_{{\rm H}({\rm curl};\Omega)}:=\|v\|^2_{L^2(\Omega)}+\|{\rm curl}\, v\|^2_{L^2(\Omega)}\, ,
\end{align}
of functions $u$ or vector fields $v$, respectively. Furthermore, we introduce their closed subspaces $H_0^1(\Omega)$, and ${\rm H}_0({\rm curl};\Omega)$ as completion under the respective graph norms of the scalar valued space $C_0^\infty(\Omega)$. Roughly speaking, $H_0^1(\Omega)$ is the subspace of functions $u\in H^1(\Omega)$ which are
zero on $\partial \Omega$, while ${\rm H}_0({\rm curl};\Omega)$ is the subspace of vectors $v\in{\rm H}({\rm curl};\Omega)$ which are normal at $\partial \Omega$ (see \cite{NeffPaulyWitsch,NPW2,NPW3}). For vector fields $v$ with components in ${\rm H}^{1}(\Omega)$ and tensor fields $P$ with rows in ${\rm H}({\rm curl}\,; \Omega)$, i.e.,
\begin{align}
v=\left(
  \begin{array}{c}
    v_1 \\
    v_2 \\
    v_3 \\
  \end{array}
\right)\, , v_i\in {\rm H}^{1}(\Omega),
\ \quad
P=\left(
  \begin{array}{c}
    P_1^T \\
    P_2^T \\
    P_3^T \\
  \end{array}
\right)\, \quad P_i\in {\rm H}({\rm curl}\,; \Omega)\,
\end{align}
we define
\begin{align}
 \nabla\,v:=\left(
  \begin{array}{c}
   {\rm grad}^T\,  v_1 \\
    {\rm grad}^T\, v_2 \\
    {\rm grad}^T\, v_3 \\
  \end{array}
\right)\, ,
\ \ \ \ {\rm Curl}\,P:=\left(
  \begin{array}{c}
   {\rm curl}^T\, P_1 \\
    {\rm curl}^T\,P_2 \\
    {\rm curl}^T\,P_3 \\
  \end{array}
\right)\,.
\end{align}

We note that $v$ is a vector field, whereas $P$, ${\rm Curl}\, P$ and ${\rm Grad}\, v$ are second order tensor fields. The corresponding Sobolev spaces will be denoted by
$
{\rm H}^1(\Omega) \ \ \text{and}\  \ {\rm H}({\rm Curl}\,; \Omega)\, .
$
We recall that  for a  fourth order tensor $\mathbb{C}$ and $X\in \mathbb{R}^{3\times 3}$, we have $\C. X\in \mathbb{R}^{3\times 3}$, $\C^T. X\in \mathbb{R}^{3\times 3}$ with the components
\begin{align}
(\C.X)_{ij}=\sum\limits_{k=1}^{3}\sum\limits_{l=1}^{3}\mathbb{C}_{ijkl}\,X_{kl}\, , \qquad (\C^T.X)_{kl}=\sum\limits_{i=1}^{3}\sum\limits_{j=1}^{3}\mathbb{C}_{ijkl}\,X_{ij}\, ,
\end{align}
while if $\mathbb{L}$ a sixth order tensor, then
\begin{align}
\mathbb{L}. Z\in \mathbb{R}^{3\times 3\times 3}\,\ \ \ \text{for all}\ \ Z\in \mathbb{R}^{3\times 3\times 3}, \ \ \ (\mathbb{L}. Z)_{ijk}=\sum\limits_{m=1}^{3}\sum\limits_{n=1}^{3}\sum\limits_{p=1}^{3}\mathbb{L}_{ijkmnp}\,Z_{mnp}\, .
\end{align}

In \cite{NeffPaulyWitsch,NPW2,NPW3}, for tensor fields $P\in {\rm H}({\rm Curl}\, ; \Omega)$ the following seminorm $|||\cdot|||$ is defined
\begin{align}
|||P|||^2:=\|\sym P\|^2_{L^2(\Omega)}+\|{\rm Curl}\, P\|^2_{L^2(\Omega)}\, .
\end{align}

From \cite{NeffPaulyWitsch,NPW2,NPW3} we have the following result:
\begin{theorem}
There exists a constant $\hat{c}$ such that
\begin{align}\label{Neff1}
\| P\|_{L^2(\Omega)}\leq \hat{c}\,|||P|||\, ,
\end{align}
for all $P\in {\rm H}({\rm Curl}\, ; \Omega)$ with vanishing restricted tangential trace on $\partial \Omega$, i.e. $P.\tau=0$ on $\partial \Omega$ .
\end{theorem}

Moreover, we have
\begin{theorem}\label{wdn}
On ${\rm H}_0({\rm Curl}\, ; \Omega)$  the norms $\|\cdot\|_{{\rm H}({\rm Curl}\, ; \Omega)}$ and $|||\cdot|||$ are equivalent. In particular,
$|||\cdot|||$ is a norm on ${\rm H}_0({\rm Curl}\, ; \Omega)$ and there exists a positive constant $c$, such that
\begin{align}\label{Neff2}
c\,\|P\|_{{\rm H}({\rm Curl}\, ; \Omega)}\leq |||P|||\, ,
\end{align}
for all $P\in {\rm H}_0({\rm Curl}\, ; \Omega)$.
\end{theorem}

Moreover,  in a forthcoming paper \cite{BNPS2}  the following results are proved:

\begin{theorem}\label{BNPS2}
There exists a positive constant $C_{DD}$, only depending on $\Omega$, such that for all $P\in{\rm H}_0({\rm Curl}\, ; \Omega)$ the following estimate holds:
\begin{align}\label{BNPSe2}
\| \Curl P\|_{L^2(\Omega)}\leq C_{DD}\,\|\dev  \Curl P\|_{L^2(\Omega)}\, .
\end{align}
\end{theorem}

\begin{theorem}\label{BNPS}
There exists a positive constant $C_{DSDC}$, only depending on $\Omega$, such that for all $P\in{\rm H}_0({\rm Curl}\, ; \Omega)$ the following estimate holds:
\begin{align}\label{BNPSe}
\| P\|_{L^2(\Omega)}\leq C_{DSDC}\,(\|\dev \sym P\|^2_{L^2(\Omega)}+\|\dev \Curl P\|^2_{L^2(\Omega)})\, .
\end{align}
\end{theorem}

\begin{corollary}\label{BNPS3}
For all $P\in{\rm H}_0({\rm Curl}\, ; \Omega)$ the following estimate holds:
\begin{align}\label{BNPSe3}
\| P\|_{L^2(\Omega)}+\| \Curl P\|_{L^2(\Omega)}\leq (C_{DSDC} +C_{DD})\,(\|\dev \sym P\|^2_{L^2(\Omega)}+\|\dev \Curl P\|^2_{L^2(\Omega)})\, .
\end{align}
\end{corollary}

\begin{theorem}\label{BNPS4}
There exists a positive constant $C_{DSG}$, only depending on $\Omega$, such that for all $u\in{\rm H}_0^1(\Omega)$ the following estimate holds:
\begin{align}\label{BNPSe4}
\| \nabla u\|_{L^2(\Omega)}\leq C_{DSG}\,\|\dev\sym \nabla u\|_{L^2(\Omega)}\, .
\end{align}
\end{theorem}

The estimates given by the above theorems will be essential in the study of our relaxed linear micromorphic elasticity model.

\subsection{Formulation of the static problem}

We consider  a relaxed version of the classical micromorphic model with  symmetric force stress $\sigma$. The relaxed model is a subset of the classical micromorphic  model in which we allow the usual elasticity tensors \cite{Eringen99} to become positive-semidefinite only \cite{NeffGhibaMicroModel}. Moreover, the number of constitutive coefficients is drastically reduced  with respect to the classical Mindlin-Eringen micromorphic elasticity model.   To be more specific, let us recall that the elastic free energy from  the classical Mindlin-Eringen micromorphic elasticity model can be written as
\begin{align}
2\, &\widehat{\mathcal{E}}(e,\varepsilon_p,\gamma)=\langle \widehat{\mathbb{C}}.\,(\nabla u-P),(\nabla u-P)\rangle
+\langle {\mathbb{H}}. \, \sym P,\sym P\rangle+
\langle \widehat{\mathbb{L}}.\,\nabla P,\nabla P\rangle\notag\\\ \ \ \ &\quad\quad\quad\quad\quad\quad+
2\langle \widehat{\mathbb{E}}.\,  \sym P ,(\nabla u-P)\rangle+
2\langle \widehat{\mathbb{F}}.\, \nabla P,(\nabla u-P)\rangle+
2\langle \widehat{\mathbb{G}}.\, \nabla P, \sym P \rangle\, ,
\end{align}
where again  $u$ is the displacement and $P$ is the
micro-distortion. The constitutive coefficients are such that
\begin{align}&\widehat{\mathbb{C}}:\mathbb{R}^{3\times 3}\rightarrow \mathbb{R}^{3\times 3},\quad\quad \quad \mathbb{H}:\Sym(3)\rightarrow\Sym(3),\quad\quad \quad \widehat{\mathbb{L}}:\mathbb{R}^{3\times 3\times 3}\rightarrow \mathbb{R}^{3\times 3\times 3}, \\ &\widehat{\mathbb{E}}: \Sym(3) \rightarrow \mathbb{R}^{3\times 3} , \quad\quad \quad \widehat{\mathbb{F}}:\mathbb{R}^{3\times 3}\rightarrow \mathbb{R}^{3\times 3}, \quad\qquad \widehat{\mathbb{G}}: \mathbb{R}^{3\times 3\times 3} \rightarrow\Sym(3), \quad\quad \quad \notag\,
\end{align}
and the classical constitutive variables are
\begin{align}
e:=\nabla u-P, \quad \quad \quad \varepsilon_p:=\sym P, \quad \quad \quad \gamma:=\nabla P \notag.
\end{align}

Our new set of {\it independent constitutive variables} for the relaxed micromorphic model is now, however,
\begin{align}\label{nsc}
\varepsilon_e:=\sym(\nabla u-P), \quad \quad \quad \varepsilon_p:=\sym P, \quad \quad \quad \alpha:=-\Curl P.
\end{align}

The system  of partial differential equations  which corresponds to this special linear anisotropic micromorphic continuum is derived from the following free energy
\begin{align}\label{energyourrel}
\quad 2\,\mathcal{E}(\varepsilon_e,\varepsilon_p,\alpha)&=\langle \C.\, \varepsilon_e, \varepsilon_e\rangle
+ \langle \H.\,\varepsilon_p, \varepsilon_p\rangle+ \langle \L.\, \alpha, \alpha\rangle\\
&=\underbrace{\langle \C.\, \sym(\nabla u-P), \sym(\nabla u-P)\rangle}_{\text{elastic energy}}
+ \underbrace{\langle \H.\,\sym P, \sym P\rangle}_{\text{microstrain self-energy}}+ \underbrace{\langle \L.\, \Curl P, \Curl P\rangle}_{\text{dislocation energy}},\notag
\\\notag
&\hspace{-1.7cm}\sigma=D_{\varepsilon_e} \, \mathcal{E}(\varepsilon_e,\varepsilon_p,\alpha)\in \Sym(3),\quad \quad
s=D_{\varepsilon_p} \, \mathcal{E}(\varepsilon_e,\varepsilon_p,\alpha),\in \Sym(3),\quad \quad
m=D_{\alpha} \, \mathcal{E}(\varepsilon_e,\varepsilon_p,\alpha)\in \mathbb{R}^{3\times3},
\end{align}
where  $\C\!:\!\Omega\rightarrow L(\mathbb{R}^{3 \times 3},\mathbb{R}^{3 \times 3})$, $\L\!:\!\Omega\rightarrow L(\mathbb{R}^{3 \times 3},\mathbb{R}^{3 \times 3})$ and $\H\!:\!\Omega\rightarrow L(\mathbb{R}^{3 \times 3},\mathbb{R}^{3 \times 3})$
are fourth order positive definite elasticity tensors,  and functions of class $C^1(\Omega)$. For the rest of the paper we assume  that the constitutive tensors
\begin{align}
&\mathbb{C}:\Sym(3)\rightarrow \Sym(3),\quad\quad \quad \mathbb{H}:\Sym(3)\rightarrow \Sym(3),\quad\quad \quad
\mathbb{L}_c:\mathbb{R}^{3\times 3}\rightarrow \mathbb{R}^{3\times 3}.
\end{align}
have the following symmetries\footnote{Minor symmetries means $\C_{ijrs}=\C_{jirs}$ and  $\C_{ijrs}=\C_{ijsr}$, while major symmetry asks $\C_{ijrs}=\C_{rsij}$. In other words, the first set of minor symmetries  $\C_{ijrs}=\C_{jirs}$ implies $\mathbb{C}:\mathbb{R}^{3\times3}\rightarrow \Sym(3)$, while from the second set of minor symmetries $\C_{ijrs}=\C_{ijsr}$ it follows $\mathbb{C}:\Sym(3)\rightarrow\mathbb{R}^{3\times3}$. Hence, together, the minor symmetries imply  $\mathbb{C}:\Sym(3)\rightarrow \Sym(3)$. The major symmetries $\C_{ijrs}=\C_{rsij}$ are enough to have $\langle \mathbb{C}.\, X,X \rangle=\langle X,\mathbb{C}.\,X \rangle$ for all $X\in\mathbb{R}^{3\times3}$.}
\begin{align}\label{simetries}
\C_{ijrs}=\C_{rsij}=\C_{jirs},\quad\quad\quad \quad \H_{ijrs}=\H_{rsij}=\H_{jirs} \qquad &\text{(minor+ major symmetries)},\\ \quad\quad\quad \quad {(\L)}_{ijrs}={(\L)}_{rsij}\qquad \qquad &\text{(only major symmetries)}\,.\notag
\end{align}

The comparison of the relaxed model with the classical Mindlin-Eringen free energy \cite{Eringen99} is then achieved through observing that
\begin{align}
\langle \widehat{\mathbb{C}}.X,X\rangle_{\mathbb{R}^{3\times 3}}&:=\langle \mathbb{C}.\sym X,\sym X\rangle_{\mathbb{R}^{3\times 3}},\notag\\\notag
\langle \widehat{\mathbb{L}}.\nabla P,\nabla P\rangle_{\mathbb{R}^{3\times 3\times 3}}&:=\langle \mathbb{L}_c.\Curl P,\Curl P\rangle_{\mathbb{R}^{3\times 3}}
\end{align}
define only {\it positive semi-definite tensors $\widehat{\mathbb{C}}$ and $\widehat{\mathbb{L}}$} when $\mathbb{C}$ and $\mathbb{L}_c$ acting on linear subspaces of $\gl(3)\cong \mathbb{R}^{3\times3}$, are assumed to be {\it strictly  positive definite tensors }.

We assume that the  fourth order elasticity tensors $\C$, $\L$ and $\H$ are positive definite.
Then, there are  positive numbers ${c_M}$, ${c_m}$ (the maximum and minimum elastic moduli for $\C$), ${(L_c)}_M$, ${(L_c)}_m$ (the maximum and minimum moduli for $\L$) and $h_M$, $h_m$ (the maximum and minimum moduli for $\H$) such that
\begin{align}\label{posdef}
\begin{array}{llll}
{c_m}\|X \|^2&\leq \langle\,\C. X,X\rangle&\leq {c_M}\|X \|^2\,\quad  \ \ &\text{for all }\ \ X\in\Sym(3),
\\
{(L_c)}_m\|X \|^2&\leq \langle \L. X,X\rangle&\leq {(L_c)}_M\| X\|^2\,\quad \ \ &\text{for all }\ \ X\in\mathbb{R}^{3\times 3},
\\
h_m\|X \|^2&\leq \langle \H. X,X\rangle&\leq h_M\| X\|^2\, \quad \ \ &\text{for all }\ \ X\in\Sym(3).
\end{array}
\end{align}
Further we  assume, without loss of generality, that ${c_M}$, ${c_m}$, ${(L_c)}_M$, ${(L_c)}_m$, $h_M$ and $h_m$  are constants.

\begin{remark}Since $P$ is determined in  $ {\rm H}({\rm Curl}\,; \Omega)$ in our relaxed model
the only possible description of boundary values is in terms of tangential traces
$P.\tau$. This follows from the standard theory of the  ${\rm H}({\rm Curl}\,; \Omega)$-space {\rm \cite{Raviart79}}.
\end{remark}

In the absence of time dependence, the basic equations \cite{NeffGhibaMicroModel} reduce to the following
system of partial differential equations (the Euler-Lagrange equations corresponding to \eqref{energyourrel})
\begin{align}\label{eqrelax}
0&=\dvg[ \C. \sym(\nabla u-P)]+f\, , \quad\quad \quad\quad\quad\quad\quad\quad\quad\quad\quad\quad\quad \ \ \ \ \  \text{balance of forces}, \\\notag
0&=- \crl[ \L.\crl\, P]+\C. \sym (\nabla u-P)-\H. \sym P+M, \quad\quad \ \  \text{balance of moment stresses},\, \ \ \
\end{align}
in $  \Omega$.
Consistently with our previous remarks, we consider the weaker (compared  to the classical) boundary conditions
\begin{align} \label{bc}
{u}({x})=0, \ \ \  \text{and the {\it tangential condition}}  \quad {P}({x}).\,\tau(x) =0 \ \ \
\ {x}\in\partial \Omega,
\end{align}%
for all tangential vectors $\tau$ at $\partial \Omega$.  In the following we suppose that the body loads satisfy the following regularity conditions
 \begin{equation}
 f,\, M\in L^2(\Omega) \label{sl}.
 \end{equation}

Since our new approach, in marked contrast to classical asymmetric micromorphic models, features
a {\it symmetric Cauchy stress tensor} $\sigma=\C. \sym(\nabla u-P)$, the linear Cosserat approach (\cite{Neff_ZAMM05,Neff_Chelminski08,Neff_JeongMMS08,Neff_Jeong_ZAMP08,Neff_Muench_simple_shear09}: $\mu_c>0$) is excluded here (see \cite{Neff_Chelminski07_disloc,NN-SIAM12,NN13,Ebobisse_Neff09,neffplastic} for further discussions).

 In contrast with  the classical 7+11 parameters of the isotropic Mindlin and Eringen model \cite{Mindlin64,Eringen64,EringenSuhubi2}, we have altogether only  seven parameters  $\mu_e,\lambda_e,  \mu_h, \lambda_h,\alpha_1, \alpha_2$, $\alpha_3$.  For isotropic materials, our system \label{eq} specializes to
\begin{align}\label{eqiso}
0&=\dvg\,\sigma+f\, ,\quad \quad
0=-\Curl m+\sigma-s+M\,  \ \ \ \text {in}\ \ \  \Omega,
\end{align}
where
\begin{align}\label{ceisor}
\sigma&= 2\mu_e \sym(\nabla u-P)+\lambda_e \tr(\nabla u-P){\cdp} \id,\notag\\
m&=\alpha_1 \dev\sym \Curl P+\alpha_2 \skew \Curl P +\alpha_3\, \tr(\Curl P){\cdp} \id,\\
s&=2\mu_h \sym P+\lambda_h \tr (P){\cdp} \id\,. \notag
\end{align}
Thus, for isotropic  elastic materials we obtain the complete system of linear partial differential equations in terms of the kinematical unknowns $u$ and $P$
\begin{align}\label{eqisoup}
0&=\dvg[2\mu_e \sym(\nabla u-P)+\lambda_e \tr(\nabla u-P){\cdp} \id]+f\, ,\\\notag
0&=-\Curl [\alpha_1 \dev\sym \Curl P+\alpha_2 \skew \Curl P +\alpha_3\, \tr(\Curl P){\cdp} \id]\\&\quad\ +2\mu_e \sym(\nabla u-P)+\lambda_e \tr(\nabla u-P){\cdp} \id-2\mu_h \sym P-\lambda_h \tr (P){\cdp} \id+M\,  \ \ \ \text {in}\ \ \  \Omega.\notag
\end{align}

In this model, {\it the asymmetric parts of $P$}, which are not suppressed,  are entirely due only to {\it moment stresses} and {\it applied body moments}. In this sense, the {\it macroscopic} and {\it microscopic scales} are fully {\it separated}.

The positive definiteness required for  the tensors $\mathbb{C}$, $\mathbb{H}$ and $\mathbb{L}_c$ implies for an isotropic material the following restriction upon the parameters $\mu_e,\lambda_e, \mu_h, \lambda_h, \alpha_1, \alpha_2$ and $\alpha_3$
\begin{align}\label{condpara}
\mu_e>0,\quad\quad  2\mu_e+3\lambda_e>0, \quad\quad  \mu_h>0, \quad\quad  2\mu_h+3\lambda_h>0, \quad\quad  \alpha_1>0,
\quad\quad  \alpha_2>0, \quad\quad  \alpha_3>0.
\end{align}
Therefore, positive definiteness for our isotropic model does not involve extra nonlinear side conditions between constitutive coefficients \cite{Eringen99,Smith}.  For the mathematical treatment of the linear relaxed model there arises the need for new integral type inequalities which we have presented in the previous subsection. Using the new results established by Neff, Pauly and  Witsch \cite{NeffPaulyWitsch,NPW2,NPW3} and by Bauer, Neff, Pauly and Starke \cite{BNPS2,BNPS3} we are now able to manage also  energies depending on the dislocation energy and  having symmetric Cauchy stresses.

If, in order to describe the mechanical behavior of a wider range of microstructured materials, we add the anti-symmetric term  $2\mu_c\skew(\nabla u-P)$ in the expression of the Cauchy stress tensor $\sigma$, where $\mu_c\geq0$ is the {\it Cosserat couple modulus},  then our analysis works also for $\mu_c\geq0$ . The model in which $\mu_c>0$ is the isotropic Eringen-Claus  model for dislocation dynamics \cite{Eringen_Claus69,EringenClaus,Eringen_Claus71} and it is, in fact, derived from the following free energy
\begin{align}\label{XXXX}
\mathcal{E}(e,\varepsilon_p,\alpha)&=\mu_e \|\sym (\nabla u-P)\|^2+\mu_c\|\skew(\nabla u-P)\|^2+ \frac{\lambda_e}{2}\, [\tr(\nabla u-P)]^2+\mu_h \|\sym  P\|^2+ \frac{\lambda_h}{2} [\tr\,(P)]^2\notag\\&
 \quad \quad  +\frac{\alpha_1}{2}\| \dev\sym \Curl P\|^2 +\frac{\alpha_2}{2}\| \skew \Curl P\|+ \frac{\alpha_3}{2}\, \tr(\Curl P)^2.
 \end{align}
  For $\mu_c>0$ and if the other inequalities \eqref{condpara} are satisfied, the existence and uniqueness follow along the well known  classical lines. There is no need for any new integral inequality. To the sake of simplicity, we only present in this paper well-posedness results for the relaxed model $\mu_c=0$. These results, however,  still hold for  $\mu_c>0$ and can be easily generalized with some additional calculations. Moreover, the results established in our paper can be easily extended to theories which include electromagnetic and thermal interactions \cite{GGI11,GalesEJMA12,Maugin13}.

\section{Existence of the solution}\label{existences}\setcounter{equation}{0}

In this section we  establish an existence
theorem for the solution of the boundary value problem $(\mathcal{P})$ defined by \eqref{eqrelax} and \eqref{bc}. To this aim,  we will rewrite the  boundary value problem
$({\mathcal{P}})$  in a weak form in a Hilbert space. The suitable Hilbert space for the equilibrium problem in the relaxed model is
\begin{equation}
\mathcal{X}\,{=}\,\big\{\,w=(u,P)\,|\,\ u{\in}\,{H}^1_0(\Omega),\quad P{\in}\, H_0(\Curl; \Omega)\big\}.
\end{equation}
According to Theorem \ref{wdn},  on $\mathcal{X}$ we have the following norm
\begin{equation}
||| w |||_\mathcal{X}=\left(\|u\|_{H_0^1(\Omega)}^2+||| P |||^2 \right)^{\frac{1}{2}},
\end{equation}
which is equivalent with the usual norm on $\mathcal{X}$
\begin{equation}
\| w \|_\mathcal{X}=\left(\|u\|_{H_0^1(\Omega)}^2+\|P\|_{{\rm H}({\rm Curl}\, ; \Omega)}^2 \right)^{\frac{1}{2}}.
\end{equation}

On $\mathcal{X}$ we define the bilinear form
\begin{align}\label{proscalar}
 (w_1,w_2)=\dd\int _\Omega\biggl( \langle \C. \sym(\nabla u_1-P_1), \sym(\nabla u_2-P_2)\rangle
+ \langle \H.\sym P_1, \sym P_2\rangle+ \langle \L. \Curl P_1, \Curl P_2\rangle\biggr)dv,\notag
\end{align}
where
$w_1=(u_1,P_1)\in\mathcal{X}$ and $w_2=(u_2,P_2)\in\mathcal{X}$.
From \cite{NeffGhibaMicroModel} we have  a first algebraic estimate:
\begin{lemma}\label{lemaii} If \, $\C$ and $\H$ satisfy the relations \eqref{simetries}$_{1,2}$ and \eqref{posdef}$_{1,3}$,  then there is a positive constant $a_1$ such that
\begin{align}
a_1(\| \sym \nabla u\|^2+\|\sym P\|^2)\leq \langle \C. \sym(\nabla u-P), \sym(\nabla u-P)\rangle+ \langle \H.\sym P, \sym P \rangle
\end{align}
for all $u\in H^1(\Omega)$ and $P\in H(\Curl;\Omega)$.
\end{lemma}

Let us define the linear operator  $l:\mathcal{X}\rightarrow\mathbb{R}$, describing the influence of external loads,
\begin{equation}\label{ling1g2}
l(\widetilde{w})=\int_\Omega (\langle f, \widetilde{u}\rangle+\langle M, \widetilde{P}\rangle )dv \ \ \ \text{for all}\ \ \ \widetilde{w}\in\mathcal{X}.
\end{equation}

We say that $w$ is a weak solution of the problem $(\mathcal{P})$ if and only if
\begin{equation}\label{wf}
(w,\widetilde{w})=l(\widetilde{w})  \ \  \text{ for all } \ \   \widetilde{w}\in \mathcal{X}.
\end{equation}

\begin{proposition}\label{lemaceg} If the constitutive coefficients satisfy the relations \eqref{simetries}, then a classical solution $w=(u,P)\in\mathcal{X}$
 of the problem $(\mathcal{P})$ is also a weak solution.
\end{proposition}
\textit{Proof.} First of all, let us recall the identities
\begin{align}
{\rm div} (\psi \, a)&=\langle a, {\rm grad} \, \psi\rangle+\psi\, {\rm div}\,  a\, ,\\
{\rm div}\,  (a\times b)&=\langle b, \curl \, a\rangle-\langle a, \curl\,  b\rangle\, ,\notag
\end{align}
for all $C^1$-functions $\psi:\Omega\rightarrow \mathbb{R}$ and $a,b:\Omega\rightarrow \mathbb{R}^{3}$, where $\times$ is the cross product. Hence
\begin{align}\label{formule}
&{\rm div} (\varphi_i Q_i)=\langle Q_i, \nabla \, \varphi_i\rangle+\varphi_i\, {\rm div}\,  Q_i,\,\quad  \ \ \text{not summed} ,\\
&{\rm div}\,  (R_i\times S_i)=\langle S_i, \curl \, R_i\rangle-\langle R_i, \curl\,  S_i\rangle,\, \quad  \ \ \text{not summed} ,\notag
\end{align}
for all $C^1$-functions $\varphi_i:\Omega\rightarrow \mathbb{R}$ and $Q_i,R_i,S_i:\Omega\rightarrow \mathbb{R}^{3}$, where $\varphi_i$ are the components of the vector $\varphi$ and $Q_i, R_i,S_i$ are the rows of the matrix $Q$, $R$ and $S$, respectively. We choose
\begin{align}
\varphi=u,\quad\quad Q=\C. \sym(\nabla u-P)
\end{align}
and we obtain
\begin{align}
&{\rm div} (u_{i} [\C. \sym(\nabla u-P)]_i)=\langle [\C. \sym(\nabla u-P)]_i, \nabla \, u_{i}\rangle+u_{i}\, {\rm div}\,  [\C. \sym(\nabla u-P)]_i,\,
 \ \ \text{not summed}\,.
\end{align}
This leads to
\begin{align}
&\sum\limits_{i=1}^3u_{i}\, {\rm div}\,  [\C. \sym(\nabla u-P)]_i=\sum\limits_{i=1}^3{\rm div} (u_{i} [\C. \sym(\nabla u-P)]_i)-\sum\limits_{i=1}^3\langle [\C.\sym(\nabla u-P)]_i, \nabla \, u_{i}\rangle\, .
\end{align}
Thus
\begin{align}\label{fdiv}
&\langle  {\rm Div}\,  [\C. \sym(\nabla u-P)], u\rangle=\sum\limits_{i=1}^3{\rm div} (u_{i} [\C. \sym(\nabla u-P)]_i)-\langle \C. \sym(\nabla u-P), \sym \nabla u\rangle\, .
\end{align}
If we take in \eqref{formule}
\begin{align}
R_i=[\L. \Curl P]_i, \quad\quad S_i=P_i\, ,
\end{align}
we have
\begin{align}
&\sum\limits_{i=1}^3{\rm div}\,  ([\L.\curl P]_i\times P_{i})=\sum\limits_{i=1}^3\langle P_{i}, \curl \, [\L. \Curl P]_i\rangle-\sum\limits_{i=1}^3\langle [\L. \Curl P]_i, \curl\,  P_{i}\rangle\, .\notag
\end{align}
Hence, we obtain
\begin{align}\label{fcurl}
&\langle P, \Curl \, (\L. (\Curl \,  P))\rangle=\sum\limits_{i=1}^3{\rm div}\,  ([\L. \Curl P]_i\times P_{i})+\langle \L. \Curl\, P, \Curl\, P\rangle\, .
\end{align}
Using \eqref{eqrelax}, \eqref{fdiv} and \eqref{fcurl} we have
\begin{align}
0=&\langle\dvg  (\C. \sym(\nabla u-P)), u\rangle+\langle f, u\rangle\, \\\notag
&-\langle\crl(\L. \crl( P)), P\rangle+\langle \C. \sym (\nabla u-P), P\rangle-\langle \H\, \sym P,  P\rangle+\langle M,  P\rangle\,\\\notag
=&-\langle \C. \sym(\nabla u-P), \sym(\nabla u-P)\rangle-\langle \L. \Curl\, P, \Curl\, P\rangle-\langle \H. \sym P,  \sym P\rangle\, \\\notag
&+\sum\limits_{i=1}^3{\rm div} (u_{i} [\C. \sym(\nabla u-P)]_i)+\sum\limits_{i=1}^3{\rm div}\,  ( P_{i}\times [\L. \Curl P]_i)+\langle f, u\rangle+\langle M,  P\rangle\,.\notag
\end{align}
Thus, we get
\begin{align}
&\int_\Omega ( \langle \C. \sym(\nabla u-P), \sym(\nabla u-P)\rangle
+ \langle \H.\sym P, \sym P\rangle+ \langle \L. \Curl P, \Curl P\rangle)dv\,  \\\notag
&=\int_{\partial \Omega} (\sum\limits_{i=1}^3 \langle [\C. \sym(\nabla u-P)]_iu_{i},n\rangle+\sum\limits_{i=1}^3\langle P_{i}\times [\L. \Curl P]_i,n\rangle)\, da+\int_\Omega (\langle f, u\rangle+\langle M,  P\rangle)\,dv,\notag
\end{align}
where $n$ is the  unit outward normal vector at the surface  $\partial  \Omega$. Therefore, using  the  boundary conditions $u=0$ and $P.\,\tau=0$ on $\partial \Omega$,  every classical solution satisfies \eqref{wf} and the proof is complete.\hfill $\Box$

\begin{theorem}
Assume that
\begin{itemize}
\item[i)] the constitutive coefficients satisfy the symmetry relations  \eqref{simetries} and  the inequalities \eqref{posdef};
\item[ii)] the loads satisfy the regularity conditions {\rm (\ref{sl})}.
\end{itemize}
Then there exists one and only one solution of the problem {\rm (\ref{wf})}.
\end{theorem}
{\it Proof.} The Cauchy-Schwarz inequality leads  to
\begin{align}
({w},\widetilde{{w}})\leq\dd\Bigg[&\int _\Omega\biggl(\langle \C. \sym(\nabla u-P), \sym(\nabla {u}-{P})\rangle\, +\langle \H. \sym P, \sym {P}\rangle+\langle \L. \Curl\, P, \Curl\, {P}\rangle\biggl) dv\Bigg]^{\frac{1}{2}}
\\\notag
&\times \Bigg[\int _\Omega\biggl(\langle \C. \sym(\nabla \widetilde{u}-\widetilde{P}), \sym(\nabla {\widetilde{u}}-{\widetilde{P}})\rangle\,+\langle \H. \sym \widetilde{P}, \sym {\widetilde{P}}\rangle+\langle \L. \Curl\, \widetilde{P}, \Curl\, {\widetilde{P}}\rangle\biggl)dv \Bigg]^{\frac{1}{2}}\, .
\end{align}

In view of \eqref{posdef} we obtain
\begin{align}
({w},\widetilde{w})\leq \dd \,C\, \Bigg[&\int _\Omega\biggl(\| \sym(\nabla u-P)\|^2+\| \sym P\|+\|\Curl\, P\|^2\biggl)dv\Bigg]^{\frac{1}{2}}
\\\notag
&\times \Bigg[\int _\Omega\biggl(\| \sym(\nabla \widetilde{u}-\widetilde{P})\|^2+\| \sym \widetilde{P}\|+\|\Curl\, \widetilde{P}\|^2\biggl)dv\Bigg]^{\frac{1}{2}}\, ,
\end{align}
where $C$ is a positive constant. Hence, we can find a positive constant $C$ such that
\begin{align}
({w},\widetilde{w})\leq \dd \,C\, \Bigg[&\int _\Omega\biggl(\| \nabla u\|^2+\|\sym P\|^2+\|\Curl\, P\|^2\biggl)dv\Bigg]^{\frac{1}{2}}
\\\notag
&\times \Bigg[\int _\Omega\biggl(\| \nabla \widetilde{u}\|^2+\|\sym \widetilde{P}\|^2+\|\Curl\, \widetilde{P}\|^2\biggl)dv\Bigg]^{\frac{1}{2}}\leq \dd \,C\, |||w|||_{\mathcal{X}}\,\,|||\widetilde{w}|||_{\mathcal{X}}\, ,
\end{align}
which means that $(\cdot,\cdot)$ is bounded. On the other hand, we have
\begin{align}
({w},{w})
=\dd\int _\Omega\biggl(&\langle \C. \sym(\nabla u-P), \sym(\nabla {u}-{P})\rangle +\langle \H. \sym P, \sym {P}\rangle+\langle \L. \Curl\, P, \Curl\, {P}\rangle\biggl)\, dv\,\notag
\end{align}
for all
${w}=({u},{P})\in \mathcal{X}$. Moreover, as a consequence of Lemma \ref{lemaii}
 and of  the assumptions \eqref{posdef} we have
\begin{align}
({w},{w})
&\geq \dd\int _\Omega\biggl(a_1\| \sym\nabla u\|^2+a_1\| \sym P\|^2+{(L_c)}_m \|\Curl P\|^2\biggl)\, dv\,
\\\notag&
\geq \dd\min\{a_1,{(L_c)}_m\}\int _\Omega\biggl(\big(\| \sym\nabla u\|^2+\|\sym P\|^2+\|\Curl P\|^2\biggl)\, dv\, .
\end{align}

Using the Korn's inequality \cite{neffKorn} and  Theorem \ref{wdn},  we deduce
\begin{align}\label{oi}
&({w},{w})
\geq \dd C\int _\Omega\biggl(\big( \| \nabla u\|^2+\|P\|^2+\|\Curl P\|^2\biggl)\, dv\,
\geq \dd C\,|||w|||^2_{\mathcal{X}}\, ,
\end{align}
where $C$ is a positive constant. Hence our bilinear form $(\cdot,\cdot)$ is coercive. Finally, the Cauchy-Schwarz inequality and the Poincar\'{e}-inequality  imply that the linear operator $l(\cdot)$ is bounded.
By the Lax-Milgram theorem it follows that (\ref{wf}) has one and only one solution. The proof is complete. \hfill$\Box$

\begin{remark} The Lax-Milgram theorem used in the proof of the previous theorem also offers a continuous dependence result on the loads $f, M$.   Moreover, the weak solution $w$  minimizes the corresponding energy functional $\dd\frac{1}{2}(w,w)-l(w)$ on $\mathcal{X}$.

\end{remark}

\section{Static problem for a further relaxed  model}
\setcounter{equation}{0}

In \cite{NeffGhibaMicroModel} a further relaxed model was proposed. This model considers an even  weaker energy expression, i.e.  it depends  only on the set of  {\it independent constitutive variables}
 \begin{align}
 \varepsilon_e=\sym(\nabla u-P),\quad\quad\quad \dev \varepsilon_p=\dev\sym P, \quad\quad\quad \dev\alpha=-\dev \Curl P.
 \end{align}
  In this model, it is neither implied that $P$ remains symmetric, nor that $P$ is trace-free, but only the trace free symmetric part of the micro-distortion $P$ and the trace-free part of the micro-dislocation tensor $\alpha$ contribute to the stored energy. The model in its general anisotropic form is then:
\begin{align}\label{eqdev}
0&={\dvg}[ \C. \sym(\nabla u-P)]+f\, ,\\\notag
0&=- {\Curl}[ \dev [\L.\dev \Curl P]]+\C. \sym (\nabla u-P)-\H. \dev \sym P+M\, \ \ \ \text {in}\ \ \  \Omega.
\end{align}
In the isotropic case the model turns into
\begin{align}\label{eqisoup2}
 0&=\dvg[2\mu_e \sym(\nabla u-P)+\lambda_e \tr(\nabla u-P){\cdp} \id]+f\, ,\\\notag
0&=-\Curl [\alpha_1 \dev\sym \Curl P+\alpha_2 \skew \Curl P ]\\&\quad\ +2\mu_e \sym(\nabla u-P)+\lambda_e \tr(\nabla u-P){\cdp} \id-2\mu_h \dev\sym P+M\,  \ \ \ \text {in}\ \ \  \Omega.\notag
\end{align}
To the system of partial differential equations of this model we adjoin the weaker boundary conditions
\begin{align} \label{bcdev}
{u}({x})=0, \ \ \ \quad \quad {P}({x}).\, \tau(x) =0  \ \ \
\ {x}\in\partial \Omega.
\end{align}%
We remark again that $P$ is not trace-free in this formulation and no projection is performed. We denote the new problem defined by the above equations and the boundary conditions \eqref{bcdev} by $(\widetilde{\mathcal{P}})$.
We observe that since $\mathbb{H}$ is positive definite on $\Sym(3)$,  in view of \eqref{posdef} we also have the estimate
\begin{align}
h_m\|\dev \sym P \|^2\leq &\langle \H. \dev \sym P,\dev \sym P\rangle\leq h_M\| \dev \sym P\|^2\,  \ \ \text{for all }\ \ P\in \mathbb{R}^{3\times3}\,.
\end{align}
Further on, we study the existence of  the solution of the problem $(\widetilde{\mathcal{P}})$. Since the method is similar with that used in Section \ref{existences} we only point out the differences which arise for our modified problem. We consider the same Hilbert space $\mathcal{X}$ as defined in Section \ref{existences} and we define the following bilinear form
\begin{align}\label{proscalar}
 ((w_1,w_2))=\dd\int _\Omega\biggl(&\langle \C. \sym(\nabla u_1-P_1), \sym(\nabla u_2-P_2)\rangle
\vspace{2mm}\notag\\&+ \langle \H.\dev\sym P_1, \dev\sym P_2\rangle+ \langle \L. \dev \Curl P_1, \dev \Curl P_2\rangle\biggr)dv,\notag
\end{align}
where
$w_1=(u_1,P_1)\in\mathcal{X}$ and $w_2=(u_2,P_2)\in\mathcal{X}$. Moreover, using a similar calculus as in the previous section and the identity
\begin{equation}\label{ABdev}
\langle \dev A,B\rangle=\langle A, \dev B\rangle, \ \  \text{for all} \ \ \ A,B\in \mathbb{R}^{3\times 3},
\end{equation}
we are able to give a weak formulation of the problem $(\mathcal{\widetilde{P}})$. We say that $w$ is a weak solution of the problem $(\mathcal{\widetilde{P}})$ if and only if
\begin{equation}\label{wf2}
((w,\widetilde{w}))=l(\widetilde{w}),  \ \  \text{ for all }  \ \   \widetilde{w}\in \mathcal{X},
\end{equation}
where $l$ is defined by \eqref{ling1g2}. In order to prove the existence of a weak solution of the problem \eqref{wf2}, let us recall that, using Theorem \ref{BNPS4}, in \cite{GhibaNeffExistence}  the following lemma was proved:
\begin{lemma}
Assume that $\C$ and $\H$ satisfy the conditions \eqref{posdef}, then  the following estimate holds true
\begin{align}
a_2\bigg(\| \nabla u\|^2_{L^2(\Omega)}&+\|\dev \sym P\|^2_{L^2(\Omega)}\bigg)\leq  \int_{\Omega}\bigg(\langle\C. \sym(\nabla u-P), \sym(\nabla u-P)\rangle+ \langle \H.\dev \sym P, \dev \sym P\rangle\bigg)dv\, ,\notag
\end{align}
for all $u\in H^1_0(\Omega)$ and $P\in H(\Curl;\Omega)$, where   $a_2$ is a positive constant.
\end{lemma}

Let us remark that in view of the above Lemma, there is a positive constant $a_3$ such that
\begin{align}
a_3\big( \| \nabla u \|^2_{L^2(\Omega)}+ \|\dev \sym P\|^2_{L^2(\Omega)}+\|\dev \Curl P \|^2_{L^2(\Omega)}\big)\leq ((w,w)),
\end{align}
where
$w=(u,v,K,P)\in \mathcal{X}$. In other words, using Corollary \ref{BNPS3} we have
\begin{align}
C\,\|w\|_{\mathcal{X}}\leq ((w,w))\, ,\ \ \text{for all}\ \  w=(u,v,K,P)\in \mathcal{X},
\end{align}
where $C$ is a positive constant. Hence $((\cdot,\cdot))$ is coercive. Moreover, the bilinear form $((\cdot,\cdot))$ is bounded, i.e. there is a positive constant $C$ such that
\begin{align}
(({w},\widetilde{w}))\leq \dd \,C\, \|w\|_{\mathcal{X}}\,\|\widetilde{w}\|_{\mathcal{X}}\, ,\ \ \text{for all}\ \  w,\widetilde{w}\in \mathcal{X}.
\end{align}

Hence, we are able to formulate the following existence result:

\begin{theorem}
Assume that
\begin{itemize}
\item[i)] the constitutive coefficients satisfy the symmetry relations  \eqref{simetries} and  the inequalities \eqref{posdef};
\item[ii)] the loads satisfy the regularity conditions {\rm (\ref{sl})}.
\end{itemize}
Then there exists one and only one solution of the problem {\rm (\ref{wf2})}.
Moreover, the weak solution $w$    minimizes the energy functional $\dd\frac{1}{2}((w,w))-l(w)$ on $\mathcal{X}$ and we have the continuous dependence of the weak solution upon the loads $f, M$.
\end{theorem}

\section{Gauge theory of dislocations}
\setcounter{equation}{0}

In this subsection we explain how we can construct
a gauge theory of dislocations and which are the relations of the constructed theory with the relaxed theory of micromorphic elastic materials, the Mindlin-Eringen/Claus-Eringen theory, and with other models of dislocations. Here we assume smooth functions, if not otherwise stated.

\subsection{Ground states in  the gauge theory of dislocations}

For dislocation gauge theory, the gauge group is the three-dimensional
translation group $T(3)$.
First of all, we may postulate a local (or soft) translation transformation for
the displacement $u$ as {\it gauge transformation}
\begin{align}\label{trg}
u^*=u+\tau\,,
\end{align}
where $\tau$ is a space-dependent (or local) translation vector.  The transformation \eqref{trg} represents the
generalization of a rigid body translation with $\tau = \text{const}$. Of course, the displacement gradient is not invariant under a local translational transformation
\begin{align}
\nabla u^*= \nabla u+\nabla \tau\,,
\end{align}
due to the second term. We require
from the corresponding energy density  to stay invariant under the internal transformation of
the displacement field. Therefore, we have to  describe the deformation using constitutive variables which are invariant under internal transformations. This justifies, in the gauge theory of dislocations, the introduction of the micro-distortion tensor $P$. In gauge theory of dislocation, the micro-distortion tensor $P$ is called  {\it the translational gauge potential}.
A starting assumption in the gauge theory of dislocation is that the micro-distortion tensor $P$ (the translational gauge potential)
possesses the following inhomogeneous transformation law
with respect to the translation group:
\begin{align}
P^*=P+\nabla \tau\,.
\end{align}
This assumption solves the invariance problem of the displacement gradient  under local translation transformation. However, together with the general invariance assumption this precludes the presence of the microstress tensor $s$ (i.e. $\H=0$). Since $P$ is a gauge potential, it transforms inhomogeneously.
Then the gauge potential  couples to the displacement field $u$ by the
{\it T(3)-gauge-covariant derivative}
\begin{align}
\label{dist3}
D^* u:=\nabla u -P=e\,,
\end{align}
i.e. the {\it elastic distortion} (relative distortion) from Mindlin-Eringen theory \cite{Eringen99} (see \cite{NeffGhibaMicroModel}). It is clear that the so called {\it T(3)-gauge-covariant derivative} is not a derivative in the common meaning.
Thus, we have redefined the elastic
distortion $ {e}$ by means of the gauge-covariant derivative
in terms of the displacement gradient (total distortion) and the plastic distortion.
We may call $ {e}$ the {\it incompatible elastic distortion}.
Now $ {e}$ is {\it gauge-invariant} under local $T(3)$-transformations
\begin{align}
\label{disl-inv-1}
e^*= {e}\,.
\end{align}
In $T(3)$-gauge theory, the Curl of the gauge potential gives rise
to an additional physical state quantity, the {\it translational field strength} (the micro-dislocation density tensor), $\alpha$,
defined by
\begin{align}
\label{disl-den}
\alpha=-\Curl P \,,
\end{align}
or in terms of $ {e}$
\begin{align}
\label{disl-den2}
\alpha=\Curl  {e}\,.
\end{align}
Thus, the translational field strength gives in a natural way the
{\it dislocation density tensor}
as state quantity.
Since $\alpha$ is a state quantity\footnote{Here, a state quantity is by definition a gauge invariant object.}, it has to be gauge-invariant:
\begin{align}
\label{disl-inv}
\alpha^*=\alpha\,.
\end{align}
In addition,
it must fulfill the so called {\it translational Bianchi identity}
\begin{align}
\label{Bianchi-iden}
\Div \alpha=0,
\end{align}
pointing out that dislocations cannot end inside the body.
Therefore, the physical state quantities (the set of constitutive variables) in the dislocation gauge theory are
\begin{align}
\label{Be}
 {e}&=\nabla u-P,\qquad
\alpha=\Curl {e}=-\Curl P\, \qquad \text{(but not only}\ P\  \text{itself)}.
\end{align}

Now some other field theoretical remarks are in order.
As pointed out by Lazar~\cite{Lazar09}
it is remarkable that the gauge-field theoretical structure
of the dislocation gauge theory may be understood using a {\it Higgs mechanism}
in the translational gauge theory.
In the translational Higgs mechanism the displacement field $u$ plays
the physical role of a Nambu-Goldstone field giving
the Proca tensor field $ {e}$, which is a physical state quantity,
a ``mass''.
Using an affine gauge approach~\cite{MAG}
it turns out that the gauge potential $P$ has the
geometrical meaning of the translational part of the generalized affine connection
and $\alpha$ is the translational part of the affine curvature
(see also~\cite{Lazar00,LH}).
A systematic investigation of conservation and balance laws in dislocation gauge theory
using Lie-point symmetries has been carried out
by~Lazar and Anastassiadis~\cite{Lazar2008,LA09b}
and Agiasofitou and Lazar~\cite{AL10}. An important result was a straightforward
definition and physical interpretation of the Peach-Koehler force analogous
to the Lorentz force in electrodynamics
since there is a lot of confusion about the physical nature
of the Peach-Koehler force in the literature.
For functionally graded materials, dislocation gauge theory was used
in~\cite{Lazar11-IJSS}.
For two-dimensional problems the gauge theory of dislocations
has been also applied to an edge dislocation in graphene~\cite{Lazar13}.

The strain energy density of the dislocation gauge theory is given by
\begin{align}
\label{W}
2\, \widetilde{\mathcal{E}}( {e},\alpha)=\langle \widehat{\sigma},e\rangle+\langle m,\alpha\rangle -2\langle \widehat{\sigma}^0, {e}\rangle\,,
\end{align}
where $\widehat{\sigma}$ denotes the {\it force-stress tensor} from the  Mindlin-Eringen theory and $m$ is the so-called
{\it pseudomoment stress tensor} \cite{Lazar2009}, i.e. the moment stress  tensor from the relaxed theory of micromorphic elastic materials \cite{NeffGhibaMicroModel} (see also \cite{Eringen_Claus69}).   The stress $\widehat{\sigma}^0$   plays
the role of a nucleation field for dislocations in the gauge theory of dislocations (a statically admissible  background field which is related to the body moment tensor  in the Eringen-Claus model  \cite{Eringen_Claus69}).
Thus, force-stress is the specific response to elastic distortion and
pseudomoment stress  (with the dimension of a moment stress  tensor)
is the specific response to dislocations.
In general, this yields: $\widehat{\sigma}=\sigma+\skew \widehat{\sigma}$, where $\sigma=\sym \widehat{\sigma}$ is the Cauchy-stress  from the relaxed theory of micromorphic elastic materials \eqref{ceisor} (see also \cite{NeffGhibaMicroModel}).
The idea of a static dislocation gauge theory is to use
three terms in the strain energy density~(\ref{W}).
The first term contains the elastic distortion field $ {e}$.
Another one proportional to the dislocation density tensor $\alpha$
having the meaning of dislocation energy density and
a term containing a background stress tensor $\widehat{\sigma}^0$, which
is needed for self-equilibrating of the dislocations. No constitutive equations are proposed for $\widehat{\sigma}^0$ which is considered to be known. Using the calculus of variations, the following field equations
can be derived when body forces $f$ are present
\begin{align}
\label{fe-gt}
0&=\dvg\,\widehat{\sigma}+f\, ,\ \ \, \quad\qquad\qquad\qquad\text{balance of forces}\\\notag
0&=-\Curl m+\widehat{\sigma}-\widehat{\sigma}^0\,,  \qquad\qquad \text{balance of dislocation stresses}\,.
\end{align}
In the balance of dislocation stresses, it can be seen that the dislocation
fields are driven by an effective stress $\widehat{\sigma}-\widehat{\sigma}^0$.
The anisotropic constitutive relations\footnote{In the relaxed micromorphic model \cite{NeffGhibaMicroModel},   for simplicity we have omitted the mixed terms. Another reason to omit the mixed terms was that for centro-symmetric materials these terms are absent and for arbitrary anisotropic materials  they would induce nonzero force-stress ${\sigma}$ for zero elastic distortion $e=\nabla u-P=0$. Moreover, we have shown \cite{NeffGhibaMicroModel}  how our energy without any mixed terms leads, in principle, to complete equations for the Cosserat model, the microstretch model and the microvoids model in dislocation format.}  are~\cite{LA09b,Lazar09}
\begin{align}
\label{CR-aniso}
\widehat{\sigma}&= \widehat{\mathbb{C}}.\, {e}+\widehat{\mathbb{B}}.\,\alpha\,,\qquad
m=\widehat{\mathbb{B}}^T.\, {e}+\L.\, \alpha \,,
\end{align}
where $\widehat{\mathbb{B}}^T$ is the fourth order tensor $\widehat{\mathbb{B}}_{klij}$,
and
\begin{align}
\widehat{\mathbb{C}}:\mathbb{R}^{3\times 3}\rightarrow \mathbb{R}^{3\times 3},
\quad\quad \quad \widehat{\mathbb{B}}:\mathbb{R}^{3\times 3}\rightarrow \mathbb{R}^{3\times 3},
\quad\quad \quad \mathbb{L}_c:\mathbb{R}^{3\times 3}\rightarrow \mathbb{R}^{3\times 3}.
\end{align}
Dimensionally, $[\mathbb{L}_c]=\ell\, [\widehat{\mathbb{B}}]=\ell^2\, [\widehat{\mathbb{C}}]$, where $\ell$ is a material length-scale parameter and, therefore,
they have the dimensions:
$[\mathbb{L}_c]=\text{force}$,
$[\widehat{\mathbb{B}}]=\text{force}/\text{length}$, and
$[\widehat{\mathbb{C}}]=\text{force}/\text{length}^2$.
Moreover, it is assumed that the material tensors satisfy the following major symmetries
\begin{align}
\label{C-sym}
\widehat{\mathbb{C}}_{ijkl}=\widehat{\mathbb{C}}_{klij}\,,\qquad\qquad ({\mathbb{L}_c})_{ijkl}=({\mathbb{L}_c})_{klij}\,.
\end{align}
Here, the tensor $\mathbb{H}$ is absent since the term $\langle \mathbb{H}.\, \sym P,\sym P\rangle$ is not translation gauge invariant. This leads to absent specific micro-stress. However, the  stress $\widehat{\sigma}_0$  is a self-equilibrating stress and incorporates the external body moment tensor $M$ from the Eringen-Claus model  \cite{Eringen_Claus69}. Hence, the strain energy density \eqref{W} of the dislocation gauge theory of anisotropic material becomes
\begin{align}
\label{Wani}
2\, \widetilde{\mathcal{E}}( {e},\alpha)=\langle \widehat{\mathbb{C}}.\, e,e\rangle+2\langle \widehat{\mathbb{B}}.\,\alpha,e\rangle +\langle \L. \, \alpha,\alpha\rangle-2\langle \widehat{\sigma}^0, {e}\rangle\,.
\end{align}
Substituting the constitutive relations~(\ref{CR-aniso}) into
(\ref{fe-gt}), we obtain
\begin{align}
\label{eom-aniso-1}
0&=\dvg[\widehat{\mathbb{C}}.\, {e}+\widehat{\mathbb{B}}.\, \alpha]+f\, ,\\\notag
\widehat{\sigma}^0&=-\Curl [\widehat{\mathbb{B}}^T.\, {e}+\L.\, \alpha]
+\widehat{\mathbb{C}}.\, {e}+\widehat{\mathbb{B}}.\,\alpha\,  \ \ \ \text {in}\ \ \  \Omega.\notag
\end{align}
On the other hand,
if we substitute Eqs.~(\ref{Be})  into (\ref{eom-aniso-1}),
we obtain the complete system of linear partial differential equations in terms of the kinematical fields $u$ and $P$ in the framework of dislocation gauge theory
\begin{align}
\label{eom-aniso}
0&=\dvg[\widehat{\mathbb{C}}.\, (\nabla u-P)+\widehat{\mathbb{B}} .\, (\Curl (\nabla u-P))]+f\, ,\\\notag
\widehat{\sigma}^0&=\Curl [\L.\, (\Curl (\nabla u-P))-\widehat{\mathbb{B}}^T.\, (\nabla u-P)]
+\widehat{\mathbb{C}}.\, (\nabla u-P)+\widehat{\mathbb{B}} .\, (\Curl (\nabla u-P))\,  \ \ \ \text {in}\ \ \  \Omega.\notag
\end{align}
Let us remark that  by applying on both sides of equation \eqref{eom-aniso}$_2$ the $\dvg$-operator, we deduce that the statically admissible  background field
$\widehat{\sigma}^0$ has to satisfy
\begin{align}\label{lambda0}
\Div \widehat{\sigma}_0+f=0 \,  \ \ \ \text {in}\ \ \  \Omega.
\end{align}
Moreover, in view of  \eqref{lambda0}, it follows that \eqref{eom-aniso}$_1$ results from \eqref{eom-aniso}$_2$. Therefore, the equations are not sufficient to find the fields $u$ and $P$ individually, but only the  elastic distortion $e=\nabla u-P$ can be determined. Thus, in terms of the elastic distortion $e$, the independent equations of the gauge theory of dislocations are
\begin{align}
\label{eom-anisoe}
\widehat{\sigma}^0&=\Curl [\L.\, (\Curl e)-\widehat{\mathbb{B}}^T.\, e]
+\widehat{\mathbb{C}}.\, e+\widehat{\mathbb{B}} .\, (\Curl e)\,  \ \ \ \text {in}\ \ \  \Omega,
\end{align}
where $\widehat{\sigma}_0$ is a solution of the problem
\begin{align}
\label{eom-anisoe1}
0&=\Div \widehat{\sigma}_0+f  \ \ \ \text {in}\ \ \  \Omega,\qquad \widehat{\sigma}_0.n=0\,  \qquad\qquad  \ \text {on}\ \ \ \partial \Omega.
\end{align}
To the system of partial differential equations of this model we adjoin the weaker tangential boundary conditions
\begin{align} \label{bcdev}
 {e}.\,\tau =0  \ \ \ \text {on}\ \ \ \partial \Omega.
\end{align}%

\subsection{Existence and uniqueness in the gauge theory of dislocations}

We  assume  in the following that
$\widehat{\sigma}^0\in L^2(\Omega)$  is known and we study the existence of the boundary value problem $(\mathcal{P}_G)$ of the gauge theory of dislocation, defined by the equations \eqref{eom-anisoe} and the boundary conditions \eqref{bcdev}. Let us consider  the following energy
\begin{align}\label{energieg}
2\, \widehat{\mathcal{E}}(e)=\langle \widehat{\C}.\, e, e\rangle
 +2\langle\widehat{\mathbb{B}}.\,\Curl\, e, e\rangle+\langle \L.\, \Curl e,  \Curl e\rangle.
 \end{align}
 In order to give a weak formulation of the boundary value problem of the gauge theory of dislocation, let us apply the equation \eqref{formule}$_2$ to the vectors
\begin{align}
R_i=[\L.\, \Curl e]_i\qquad \text{and}  \qquad K_i=[\widehat{\mathbb{B}}^T.\,  e]_i\,,
\end{align}
to deduce
\begin{align}\label{fcurlg1}
\langle e, \Curl \, (\L. \Curl \,  e)\rangle&=\sum\limits_{i=1}^3{\rm div}\,  ([\L. \Curl e]_i\times e_{i})+\langle \L. \Curl\, e, \Curl\, e\rangle\,,
\\
\langle e, \Curl \, (\widehat{\mathbb{B}}^T.\,  e))\rangle&=\sum\limits_{i=1}^3{\rm div}\,  ([\widehat{\mathbb{B}}^T.\,  e]_i\times e_{i})+\langle \widehat{\mathbb{B}}^T.\,  e, \Curl\, e\rangle\, ,\notag
\end{align}
where $e_i$ are the rows of the tensor $e\in \R^{3\times3}$.

We consider the same Hilbert space ${\rm H}_0({\rm Curl}\, ; \Omega)$  and we define the following bilinear form
\begin{align}\label{proscalarg}
 [e,\widetilde{e}]=\dd\int _\Omega\biggl(&\langle \widehat{\C}.\, e, \widetilde{e}\,\rangle
 +\langle\widehat{\mathbb{B}}.\,(\Curl\, e), \widetilde{e}\,\rangle+ \langle\widehat{\mathbb{B}}.\,(\Curl\, \widetilde{e}), e\rangle+\langle \L.\,  \Curl e,  \Curl \widetilde{e}\,\rangle\biggr)dv,
\end{align}
where
$e,\widetilde{e}\in{\rm H}_0({\rm Curl}\, ; \Omega)$. Let us define the linear operator  $l:{\rm H}_0({\rm Curl}\, ; \Omega)\rightarrow\mathbb{R}$
\begin{equation}\label{ling1g2g}
\widehat{l}(\widetilde{e})=\int_\Omega \langle \widehat{\sigma}_0, \widetilde{e}\,\rangle\, dv \ \ \ \text{for all}\ \ \ \widetilde{e}\in\mathcal{X}.
\end{equation}

We say that $e$ is a weak solution of the following boundary values  problem $(\mathcal{P}_G)$  if and only if
\begin{equation}\label{wfgg}
[e,\widetilde{e}]=\widehat{l}(\widetilde{e})  \ \  \text{ for all } \ \   \widetilde{e}\in {\rm H}_0({\rm Curl}\, ; \Omega).
\end{equation}

\begin{theorem}\label{thexgauge}
Assume that
\begin{itemize}
\item[i)] the constitutive coefficients\footnote{There are no explicit, separate assumption upon the tensors $\widehat{\mathbb{C}},\widehat{\mathbb{B}},\L$. In the admissible case $\mathbb{B}=0$, the positive definiteness of the internal energy $\widehat{\mathcal{E}}$ is equivalent with the positive definiteness of the $\widehat{\mathbb{C}},\L$. The existence results hold also true for $\widehat{\mathbb{B}}=0$.} satisfy the symmetry relations  \eqref{C-sym};
\item[ii)] there are the positive constants $c_1,c_2$ such that\,\footnote{This condition is weaker than the positive definiteness of the energy $\widehat{\mathcal{E}}(e)$ in terms of $e$ and ${\rm Curl}\, e$ and shows that the existence result may   also work for zero Cosserat couple modulus $\mu_c=0$.}
 \begin{align*}
  c_1\,\int _\Omega\bigl(&\|{\rm sym}\,e\|^2+\|\Curl\, e\|^2\bigr)dv\leq \int_\Omega\widehat{\mathcal{E}}(e)\, dv\leq c_2\,\int _\Omega\bigl(\|e\|^2+\|\Curl\, e\|^2\bigr)dv\, \qquad \forall \,e\in {\rm H}_0({\rm Curl}\, ; \Omega);
 \end{align*}
\item[iii)] the statically admissible  background field
$\widehat{\sigma}^0$  satisfies the regularity condition $\widehat{\sigma}^0\in L^2(\Omega)$.
\end{itemize}
Then there exists one and only one solution $e$ of the problem {\rm (\ref{wfgg})}.
\end{theorem}
{\it Proof.} It is simple to prove that the Cauchy-Schwarz inequality and the hypothesis ii) lead to  the boundedness  of  $[\cdot,\cdot]$. Besides this, from hypothesis ii) and using Theorem \ref{wdn}, there are the  positive constants $C_1,C_2>0$ such that
\begin{align}\label{proscalarg1}
 [e,e]\geq C_1\,\dd\int _\Omega\bigl(&\|{\rm sym}\,e\|^2+\|\Curl\, e\|^2\bigr)dv\geq C_2\, \|\, e\|^2_{{\rm H}_0({\rm Curl}\, ; \Omega)},
\end{align}
for all
$e\in {\rm H}_0({\rm Curl}\, ; \Omega)$, i.e.  $[\cdot,\cdot]$ is coercive. Finally,  the Schwarz inequality implies that the linear operator $\widehat{l}(\cdot)$ is bounded.
By the Lax-Milgram theorem it follows that (\ref{wfgg}) has one and only one solution. \hfill$\Box$

\begin{remark} The Lax-Milgram theorem used in the proof of the previous theorem also offers a continuous dependence result on the loads $f$.   Moreover, the weak solution $e$  minimizes the energy functional $\dd\frac{1}{2}[e,e]-\widehat{l}(e)$ on ${\rm H}_0({\rm Curl}\, ; \Omega)$.

\end{remark}
\subsection{The gauge theory of dislocations for isotropic materials}

For isotropic constitutive relations, Lazar~\cite{LazarJPMG02} and
Lazar and Anastassiadis~\cite{Lazar2009} have decomposed the dislocation density tensor
$\alpha$ into its ${\rm SO}(3)$-irreducible pieces (see \eqref{odesc}),  called ``the axitor", ``the tentor"
and ``the trator" parts, i.e.
\begin{align}
\label{tratoraxi}
\alpha&=\underbrace{\dev\sym \alpha}_{\alpha^{(1)}:\textrm{``tentor"}}+
\underbrace{\skew \alpha}_{\alpha^{(2)}:\textrm{``trator"}}
+\underbrace{\frac{1}{3}\,\tr (\alpha){\cdp} \id}_{\alpha^{(3)}:\textrm{``axitor"}}\, .
\end{align}
In general, the dislocation density tensor reads in matrix-form
\begin{align}
\label{DD-M}
\alpha=\left(
  \begin{array}{ccc}
    \alpha_{11}&\alpha_{12}&\alpha_{13} \\
    \alpha_{21}&\alpha_{22}&\alpha_{23} \\
    \alpha_{31}&\alpha_{32}&\alpha_{33} \\
  \end{array}
\right)\, .
\end{align}
The indices  $i$ and $j$ of $\alpha_{ij}$ determine the orientation of the
Burgers vector and the dislocation line, respectively.
Therefore, the diagonal components describe screw dislocations and
the off-diagonal components describe edge dislocations.
Substituting of ~(\ref{DD-M}) into (\ref{tratoraxi}),
the {\it axitor} reads
\begin{align}
\label{axitor}
\alpha^{(3)}=\frac{\alpha_{11}+\alpha_{22}+\alpha_{33}}{3}
\left(
\begin{array}{ccc}
    1&0&0 \\
    0&1&0 \\
    0&0&1 \\
  \end{array}
\right)\, ,
\end{align}
describing the sum of all possible screw dislocations,
the {\it trator} is given by
\begin{align}
\label{trator}
\alpha^{(2)}=\frac{1}{2}\left(
  \begin{array}{ccc}
    0&\alpha_{12}-\alpha_{21}&\alpha_{13}-\alpha_{31} \\
    \alpha_{21}-\alpha_{12}&0&\alpha_{23}-\alpha_{32} \\
    \alpha_{31}-\alpha_{13}&\alpha_{32}-\alpha_{23}&0 \\
  \end{array}
\right)\,,
\end{align}
describing ``skew-symmetric'' edge dislocations with the property
$\alpha^{(2)}=-(\alpha^{(2)})^T$,
and the {\it tentor} possesses the form
\begin{align}
\label{trenor}
\alpha^{(1)}=\frac{1}{2}\left(
  \begin{array}{ccc}
    2\,\alpha_{11}&\alpha_{12}+\alpha_{21}&\alpha_{13}+\alpha_{31} \\
    \alpha_{21}+\alpha_{12}&2\,\alpha_{22}&\alpha_{23}+\alpha_{32} \\
    \alpha_{31}+\alpha_{13}&\alpha_{32}+\alpha_{23}&2\,\alpha_{33} \\
  \end{array}
\right)\,
-\frac{\alpha_{11}+\alpha_{22}+\alpha_{33}}{3}
\left(
\begin{array}{ccc}
    1&0&0 \\
    0&1&0 \\
    0&0&1 \\
  \end{array}
\right)\,,
\end{align}
describing ``symmetric'' edge dislocations with the property
$\alpha^{(3)}=(\alpha^{(3)})^T$ and also single screw dislocations.

For the three-dimensional elastoplastic dislocation problem, \textit{screw dislocations} correspond to
\begin{align}
P=\left(
    \begin{array}{ccc}
      0 & P_{12} & P_{13} \\
      P_{21} & 0 & P_{23} \\
      P_{31} & P_{32} & 0 \\
    \end{array}
  \right)\qquad \mapsto \qquad \alpha=\Curl P=\left(
    \begin{array}{ccc}
      \alpha_{11} & 0 & 0 \\
        0& \alpha_{22}& 0 \\
      0 & 0 & \alpha_{33} \\
    \end{array}
  \right)\,,
\end{align}
while \textit{edge dislocations} correspond to
\begin{align}
P=\left(
    \begin{array}{ccc}
      P_{11} & P_{12} & P_{13}\\
      P_{21} & P_{22} & P_{23} \\
      P_{31} & P_{32} & P_{33} \\
    \end{array}
  \right)
\qquad \mapsto \qquad \alpha=\Curl P=\left(
    \begin{array}{ccc}
      0 & \alpha_{12} & \alpha_{13} \\
      \alpha_{21} & 0 & \alpha_{23} \\
      \alpha_{31} & \alpha_{32} & 0 \\
    \end{array}
  \right)\,.
\end{align}
Both situations are connected with  displacement vector $u=(u_1,u_2,u_3)$.

The isotropic constitutive relations are given by~\cite{Lazar2009}
\begin{align}
\label{CR-iso}
\widehat{\sigma}&= 2{\mu}_e  \sym  {e}+2{\mu_c} \skew  {e}+{\lambda}_e\, \tr ({e}){\cdp} \id
\,,\\
m&=\alpha_1\dev\sym \alpha+\alpha_2\skew\alpha+{\alpha_3}\,\tr(\alpha ) {\cdp}\id\,.\notag
\end{align}
The positive semi-definiteness required for the tensors $\widehat{\mathbb{C}}$, $\widehat{\mathbb{B}}$ and $\L$ implies for isotropic materials
the following restriction upon the parameters $\mu_e,\lambda_e, \mu_c, \alpha_1, \alpha_2$ and $\alpha_3$
\begin{align}
\label{condpara-gt}
\mu_e\ge0,\quad\quad  2\mu_e+3\lambda_e\ge0, \quad\quad  \mu_c\ge0,
\quad\quad  \alpha_1\ge0,
\quad\quad  \alpha_2\ge0, \quad\quad  \alpha_3\ge0\,.
\end{align}
If we put $\mu_c=0$ in (\ref{CR-iso})$_1$, the force-stress  tensor becomes
symmetric $\widehat{\sigma}=\sigma$. The static field equations
used  by Lazar and Anastassiadis \cite{Lazar2009} in the isotropic gauge theory of dislocations  read
\begin{align}\label{L}
\widehat{\sigma}^0&=\dd\Curl[\alpha_1\,\dev\sym(\Curl { {e}})+
\alpha_2\,\skew(\Curl  {e})
+{\alpha_3}\,\tr(\Curl{ {e}}){\cdp} \id]\vspace{1mm}\\\  &\qquad\dd\quad+2{\mu}_e  \sym {e}+2{\mu_c} \skew {e}+{\lambda}_e\, \tr( {e}){\cdp} \id\, ,\notag
\end{align}
or equivalently in terms of the displacement vector $u$ and plastic distortion tensor $P$
\begin{align}
\label{L2}
\widehat{\sigma}^0=&\dd-\Curl[\alpha_1\dev\sym(\Curl {P})+
\alpha_2\skew(\Curl P)
+{\alpha_3}\tr(\Curl{P}){\cdp} \id]\\\  &\qquad\quad\dd+2{\mu}_e  \sym(\nabla u-{P})+2{\mu_c} \skew(\nabla u-{P})+{\lambda}_e\, \tr(\nabla u-{P}){\cdp} \id\, ,\notag
\end{align}
where the coefficients $\alpha_1,\,\alpha_2,\, \alpha_3$  correspond to $a_1,\,a_2,\, \dd\frac{a_3}{3}$ from Lazar's original notations (see, e.g.,~\cite{Lazar2009}).
In \cite{Lazar2009} various non-singular special solutions to \eqref{L}
for screw and edge dislocations with dislocation line
in $x_3$-direction were  constructed.
A solution of a screw dislocation in a functionally graded material
within the gauge theory of dislocations was given in~\cite{Lazar11-IJSS}.
The anti-plane strain of a screw dislocation corresponds to
\begin{align}
P=\left(
    \begin{array}{ccc}
      0 & 0 &0 \\
      0 & 0 & 0 \\
      P_{31}(x_1,x_2) & P_{32}(x_1,x_2) & 0 \\
    \end{array}
  \right)\quad  \mapsto \quad \alpha=\Curl P=\left(
    \begin{array}{ccc}
      0 & 0 & 0 \\
      0 & 0 & 0 \\
      0 & 0  & \alpha_{33}(x_1,x_2)\\
    \end{array}
  \right),
\end{align}
which is connected with the following displacement vector
\begin{align}
\label{u-aps}
u=\big(0,0,u_3(x_1,x_2)\big)^T\,.
\end{align}
The plane strain problem of  edge dislocations corresponds to
\begin{align}
P=\left(
    \begin{array}{ccc}
      P_{11}(x_1,x_2) &  P_{12}(x_1,x_2) &  0\\
      P_{21}(x_1,x_2) &  P_{22}(x_1,x_2) & 0 \\
      0 & 0 & 0 \\
    \end{array}
  \right)\quad \mapsto \quad \alpha=\Curl P=\left(
    \begin{array}{ccc}
      0 & 0 & \alpha_{13}(x_1,x_2) \\
      0 & 0 & \alpha_{23}(x_1,x_2)  \\
      0 & 0 & 0 \\
    \end{array}
  \right)\, ,
\end{align}
and the corresponding displacement vector reads
\begin{align}
\label{u-ps}
u=\big(u_1(x_1,x_2),u_2(x_1,x_2),0)\big)^T\,.
\end{align}

For a screw dislocation the tentor and the axitor give a non-zero contribution
while for an edge dislocation the tentor and the trator give a non-zero contribution.
Such gauge theoretical solutions can be physically meaningful
(e.g., regularization of the stress and strain singularities,
natural dislocation core spreading making redundant the artifical cut-off radius,
and appearance of characteristic length scale parameters).

In the variational formulation, the dislocation model can be seen as an elastic (reversible) description of a material, which may respond to external loads by an elastic distortion field $ {e}$  which is not anymore a gradient (incompatible). This is not yet an irreversible  plasticity formulation, since elasticity does not change the state of the body (by definition).

\subsection{Special cases of the gauge model of dislocations}
Since the {\it Lazar-Anastassiadis gauge model of dislocations}~\cite{Lazar2009} (where the microstress is not taken into account)
with six material parameters,
$\mu_e$, $\lambda_e$, $\mu_c$, $\alpha_1$, $\alpha_2$, $\alpha_3$,
is a general gauge model for dislocations in a linear isotropic medium,
it contains some interesting special cases
based on particular assumptions on the material moduli.
Special cases are:
\begin{itemize}
\item
{\it Force stresses are symmetric}: $\mu_c=0$. This case is a particular case ($\mathbb{H}=0$, i.e. no specific microstress) of the {\it  relaxed model}.
\item
{\it Force stresses are symmetric and no axitor}: $\mu_c=0$, $\alpha_3=0$.  For this we retrieve a particular case ($\mathbb{H}=0$) of  the {\it further relaxed model}.
\item
{\it Edelen gauge model of dislocations}~\cite{KE,Edelen83,EdelenLagoudas}
with the following choice:
\begin{align}
\label{EC}
\alpha_1=\alpha_2\,,\qquad
\alpha_1=3\,\alpha_3\,,\qquad
\mu_c=0\,.
\end{align}
\item
{\it Popov-Kr\"oner gauge model of dislocations}~\cite{PK01}
with the following choice:
\begin{align}
\label{PK}
\alpha_1=\frac{3\,\mu_e (2d)^2}{24}\,,\qquad
\alpha_2=\frac{\mu_e (2d)^2}{24}\, \frac{3+\nu}{1-\nu} \,,\qquad
\alpha_3=0\,,\qquad
\mu_c=0\,,
\end{align}
where $d$ is a characteristic mesoscopic length,
and therefore
\begin{align}
\label{PK2}
\alpha_2=\frac{(3+\nu)\,\alpha_1}{3(1-\nu)}\,.
\end{align}
\item
{\it Einstein choice}~\cite{Malyshev,Lazar02,LazarJPMG02}
\begin{align}
\label{Einstein}
\alpha_1=-\alpha_2\,,\qquad
\alpha_1=-6\,\alpha_3\,,\qquad \mu_c=0\,.
\end{align}
It is called the ``Einstein choice'' since, with this choice, the dislocation energy,
$\frac{1}{2}\langle  m,\alpha\rangle$, is equivalent (up to a boundary term)
to the three-dimensional Einstein-Hilbert Lagrangian  (e.g. \cite{MAG,LazarJPMG02,LH}). Further comments regarding the Einstein choice are included in Subsection \ref{EChoice}.
\item
{\it Strain gradient-like choice}~\cite{Lazar2009} (see also~\cite{Lazar2003})
\begin{align}
\label{grad}
\alpha_2=\frac{1+\nu}{1-\nu}\,\alpha_1\,,\qquad
\alpha_1=-6\,\alpha_3\,,\qquad \mu_c=0\,.
\end{align}
The interesting feature of this choice is that the solutions of
the stress fields of screw and edge dislocations given in~\cite{GA,LM}
in the framework of strain gradient elasticity can be reproduced.
\end{itemize}
However, the Einstein choice~(\ref{Einstein})
and the choice~(\ref{grad}) do not satisfy the positivity condition~(\ref{condpara-gt}).

\subsection{Fundamental solution of force stresses
and characteristic lengths}
In general,
a {\it characteristic length} is an important dimension that defines the scale of a physical system.
For instance, in gradient elasticity
the characteristic lengths are in the range of the lattice parameters,
that is in the order $\ell \sim 10^{-10}$~m (see, e.g.,~\cite{Shodja}).
Therefore, such a theory can be used for understanding the nano-mechanical
phenomena at such length scales.
In all generalized elasticity  theories  (e.g., micropolar elasticity, gradient elasticity)
where the material tensors have different dimensions, characteristic lengths appear~(e.g.,
\cite{Eringen99,Nowacki86,Mindlin64}).
Thus, the existing characteristic length scales are given in terms of the material tensors with different dimensions.
Their structure can be seen directly in characteristic field equations and they appear explicitly in the construction of
the Green tensors (fundamental solutions) of the field equations (e.g.,~\cite{Eringen99,Nowacki86,Mindlin64}).
All fundamental solutions of linear generalized elasticity contain the corresponding
characteristic length scales as parameters.

Following Lazar and Anastassiadis~\cite{Lazar2009}, we give
the Green tensor, which is the fundamental solution of equation~(\ref{L})$_2$,
in terms of force stresses $\widehat{\sigma}$. If the Cosserat couple modulus $\mu_c>0$ \cite{Neff_Jeong_Conformal_ZAMM08,Neff_Paris_Maugin09,Neff_Jeong_IJSS09,NeffJeongFischle}, we have the inverse constitutive relation for $e$
\begin{align}
\label{CR-B}
e= \frac{\mu_c + \mu_e}{4\mu_e\,\mu_c}\,\widehat{\sigma}+
\frac{\mu_c - \mu_e}{4\mu_e\,\mu_c}\,(\widehat{\sigma})^T
- \frac{\nu}{2\mu_e (1 + \nu)}\, \tr(\widehat{\sigma}){\cdp} \id\,,
\end{align}
where the Poisson's ratio $\nu$ is expressed in terms of the Lam{\'e} coefficients $\lambda_e$ and $\mu_e$
\begin{align}
\label{}
\nu=\frac{\lambda_e}{2\,(\lambda_e + \mu_e)}\,,\qquad
\lambda_e=\frac{2\,\mu_e\,\nu}{1-2\nu}\,.
\end{align}
Using the force equilibrium condition
${\rm Div}\,\widehat{\sigma}=0$ for vanishing forces and \eqref{CR-B}, we write the equation \eqref{L} only in terms of $\sigma$
\begin{align}
\label{ME-G}
\square_{\rm G}\, \widehat{\sigma}=\widehat{\sigma}^0,
\end{align}
where $\square_{\rm G}:\R^{3\times 3}\rightarrow\R^{3\times 3}$ is a differential matrix operator  defined by
\begin{align}
(\square_{\rm G}\, \widehat{\sigma})_{ij}&=\frac{1}{4\mu_e\mu_c}
  \Big[(c_1 - c_2 + 2c_3)\,\frac{2\mu_c\nu}{1 + \nu} - 2c_3\mu_c\Big]
(\delta_{ij}\, \Delta\,\widehat{\sigma}_{ll} -\pd_i\pd_j\widehat{\sigma}_{ll})
-\big[c_1(\mu_c + \mu_e) - c_2(\mu_c-\mu_e)\big] \Delta\,\widehat{\sigma}_{ij}
\notag\\&
+ \big[c_1(\mu_c - \mu_e) - c_2(\mu_c+\mu_e)\big](
\pd_j\pd_k\widehat{\sigma}_{ki}- \Delta\,\widehat{\sigma}_{ji})
+ \big[2c_2\mu_e -c_3(\mu_c + \mu_e)\big]\,\pd_i\pd_k\widehat{\sigma}_{kj}
+4\mu_e\mu_c\,\widehat{\sigma}_{ij}\,,
\end{align}
and
\begin{align}
 \pd_i=\frac{\pd}{\pd x_i}\,,\qquad \Delta= \frac{\pd^2}{\pd x_i^2}\,, \qquad
c_1:=\frac{1}{3}\big(2\alpha_1+3\alpha_3\big)\,,\qquad
c_2:=\frac{1}{3}\big(3\alpha_3-\alpha_1\big)\,,\qquad
c_3:=\frac{1}{2}\big(\alpha_2-\alpha_1\big)\,.
\end{align}

 A fundamental solution of the
equation \eqref{ME-G}   is a matrix field $\Sigma\in\R^{3\times3}$
 which satisfies the condition \cite{Hormander}
\begin{equation}
\square_{\rm G}\, \widehat{\Sigma}=\delta({x})\cdot L\
\qquad \forall\,  x\in\R^{3},
\end{equation}
where $\delta(\cdot)$ is the Dirac delta and
$L\in\R^{3\times 3}$ has constant components. Hence, the fundamental solution is the solution of equation \eqref{ME-G} corresponding to a ``point body pseudo-moment'' or given
Dirac-delta point stress of magnitude $L$. Eventually, the fundamental solution can be written in the following form
\begin{align}
\label{T-G}
 \widehat{\Sigma}=G.\, L\,,
\end{align}
where $G: \R^{3\times 3}\rightarrow \R^{3\times 3}$ is a fourth order tensor called the Green tensor of Eq.~(\ref{ME-G}).
In this way, the three-dimensional  Green tensor of Eq.~(\ref{ME-G})
is given by (see \cite{Lazar2009})
\begin{align}
\label{GT}
G_{ijkl}&=
\frac{1}{8\pi}\bigg\{
(\delta_{ik}\delta_{jl}+\delta_{il}\delta_{jk})
\,\frac{\e^{-r/\ell_1}}{ \ell_1^2\, r}
+(\delta_{ik}\delta_{jl}-\delta_{il}\delta_{jk})
\,\frac{\e^{-r/\ell_4}}{ \ell_4^2\, r}
-(\delta_{ij}\Delta-\pd_i\pd_j)\delta_{kl}
\Big[\frac{1}{r}\big(\e^{-r/\ell_1}-\e^{-r/\ell_2}\big)\Big]\nonumber\\
&\qquad\
-(\delta_{jl}\pd_i+\delta_{il}\pd_j)\pd_k
\Big[\frac{1}{r}\big(\e^{-r/\ell_1}-\e^{-r/\ell_3}\big)\Big]
-(\delta_{jl}\pd_i-\delta_{il}\pd_j)\pd_k
\Big[\frac{1}{r}\big(\e^{-r/\ell_4}-\e^{-r/\ell_3}\big)\Big]
\bigg\}\,,
\end{align}
where $r=\sqrt{x_1^2+x_2^2+x_3^2}$ and
\begin{align}
\label{L1}
&\ell_1^2=\frac{\alpha_1}{2\,\mu_e}\,,\qquad
\ell_2^2=\frac{(1-\nu)\alpha_2}{2\,\mu_e\,(1+\nu)}\,,\qquad
\ell_3^2=\frac{(\mu_e+\mu_c)(\alpha_1+\alpha_2)}{8\,\mu_e\,\mu_c}\,,\qquad
\ell_4^2=\frac{\alpha_1+6\alpha_3}{6\,\mu_c}\,.
\end{align}
The solution~(\ref{T-G}) with (\ref{GT})
represents the three-dimensional force stress field of a point stress.
Therefore, we solved the (force) stress problem
of a concentrated body pseudo-moment
(or concentrated background stress $\widehat{\sigma}^0$)
for vanishing body forces in an unbounded material, provided the invertibility  formula \eqref{CR-B} holds true.
Using the convolution theorem, we obtain  the particular solution
of the force stress tensor $\widehat{\sigma}$
as convolution of the Green tensor
with $\widehat{\sigma}^0$
\begin{align}
\widehat{\sigma}_{ij}=G_{ijkl}*\widehat{\sigma}_{kl}^0\, ,
\end{align}
where $*$ denotes the convolution.
With the help of this equation, three-dimensional problems can be solved
for any given $\widehat{\sigma}^0$ or any body pseudo-moment tensor\footnote{Only for an unbounded domain and $\mu_c>0$.}.
From Eq.~(\ref{L1}) it can be seen that in
the Lazar-Anastassiadis dislocation model
four characteristic lengths can be defined
in terms of the six material parameters of an isotropic material,
$\mu_e$, $\lambda_e$, $\mu_c$, $\alpha_1$, $\alpha_2$, $\alpha_3$,
(e.g.,~\cite{Lazar2009}).
In addition, the lengths $\ell_1$, $\ell_2$ and $\ell_3$ fulfill the following relation
\begin{align}
\ell_3^2=\frac{\mu_e+\mu_c}{4\mu_c}\Big(\ell_1^2+\frac{1+\nu}{1-\nu}\, \ell_2^2\Big)\,.
\end{align}
Thus,  four characteristic length scales exist in the
static and isotropic gauge theory of dislocations.
The characteristic length $\ell_1$ depends on $\mu_e$ and
is similar in the form to the internal length in the couple stress theory~\cite{MT62,Mindlin63}.
Moreover, it can be seen that $\ell_2$ depends on the Poisson's ratio $\nu$.
Therefore, the length $\ell_2$ is the characteristic length of dilatation.
Because the characteristic lengths $\ell_3$ and $\ell_4$ depend on $\mu_c$ they
look like the two characteristic lengths of micropolar elasticity, namely the characteristic lengths
for bending and torsion (see, e.g.,~\cite{Nowacki86}).
The parameter of the axitor $\alpha_3$ gives only a contribution to the characteristic
length $\ell_4$.
For $\mu_c\rightarrow\infty$, only the bending length, which is the
characteristic length in the theory of couple stresses,  survives
\begin{align}
\label{L3-0}
&
\lim_{\mu_c\rightarrow\infty}\ell_3^2=\frac{\alpha_1+\alpha_2}{8\,\mu_e},
\qquad
\lim_{\mu_c\rightarrow\infty}\ell_4^2=0.
\end{align}
On the other hand, if $\mu_c\rightarrow 0,\  \mu_c\geq 0$, then
$\ell_3$ and $\ell_4$ diverge. Moreover, for $\mu_c\rightarrow 0,\  \mu_c\geq 0$
\begin{align}
\label{GT}
G_{ijkl}&\rightarrow
\frac{1}{8\pi}\bigg\{
(\delta_{ik}\delta_{jl}+\delta_{il}\delta_{jk})
\,\frac{\e^{-r/\ell_1}}{ \ell_1^2\, r}
-(\delta_{ij}\Delta-\pd_i\pd_j)\delta_{kl}
\Big[\frac{1}{r}\big(\e^{-r/\ell_1}-\e^{-r/\ell_2}\big)\Big]\nonumber\\
&\quad\qquad\
-(\delta_{jl}\pd_i+\delta_{il}\pd_j)\pd_k
\Big[\frac{1}{r}\big(\e^{-r/\ell_1}-1\big)\Big]
\bigg\}\,.
\end{align}
For symmetric force stresses, the only relevant characteristic lengths are $\ell_1$ and $\ell_2$ which are given in terms of the material parameters $\alpha_1$, $\alpha_2$, $\mu_e$ and $\nu$.
For that reason the axitor might be neglected in the case of symmetric force stress model like
in the further relaxed model. However, for $\mu_c\rightarrow0$ and $\widehat{\sigma}$ symmetric the constitutive equation \eqref{CR-iso} is not invertible. Only  the symmetric elastic strain $\sym e$ may be determined as function of $\widehat{\sigma}$. Therefore,  Lazar's approach towards special solutions via Green's function needs finally the invertibility of the force stress $\sigma$ as function of $e$. This is only possible for $\mu_c>0$. Nevertheless, our theorem \ref{thexgauge} provides existence and uniqueness for $\mu_c=0$ in bounded domains.

In the {\it Edelen choice}~(\ref{EC}),
the characteristic lengths~(\ref{L1}) become
\begin{align}
\label{L1-Ed}
\ell_1^2=\frac{\alpha_1}{2\,\mu_e}\,,\qquad
\ell_2^2=\frac{(1-\nu)\alpha_1}{2\,\mu_e(1+\nu)}\,,\qquad
\ell_3^2\quad \text{not defined}\,,\qquad
\ell_4^2\quad \text{not defined}\,.
\end{align}
The lengths $\ell_1$ and $\ell_2$ coincide with the lengths $M^{-1}$
and $N^{-1}$ introduced by Kadi\'c and Edelen~\cite{Edelen83} and  Edelen and Lagoudas~\cite{EdelenLagoudas}
in the dislocation gauge theory.
Due to $\mu_c\rightarrow 0,\ \mu_c\geq 0$ (symmetric force stresses),
$\ell_3$ and $\ell_4$ formally do not exist.

For the so-called {\it Popov-Kr\"oner choice}~(\ref{PK}) and (\ref{PK2}),
the characteristic lengths~(\ref{L1}) become
\begin{align}
\label{L1-PK}
\ell_1^2=\frac{\alpha_1}{2\,\mu_e}\,,\qquad
\ell_2^2=\frac{(3+\nu)\alpha_1}{6\,\mu_e(1+\nu)}\,,\qquad
\ell_3^2\quad \text{not defined}\,,\qquad
\ell_4^2\quad \text{not defined}\,.
\end{align}
Again due to $\mu_c=0$ (symmetric force stresses),
$\ell_3$ and $\ell_4$ formally do not exist.

For the so-called {\it Einstein choice}~(\ref{Einstein}),
the characteristic lengths~(\ref{L1}) reduce to
\begin{align}
\label{L1-HE}
\ell_1^2=\frac{\alpha_1}{2\,\mu_e}\,,\qquad
\ell_2^2=-\frac{(1-\nu)\,\alpha_1}{2\,\mu_e(1+\nu)}\,,\qquad
\ell_3^2=0\,,\qquad
\ell_4^2=0\,.
\end{align}
Thus, only two characteristic lengths survive. But the length $\ell_2$ is now imaginary.
These lengths $\ell_1$ and $\ell_2$ agree with the lengths ${\cal{M}}^{-1}$
and ${\cal{N}}^{-1}$ used by Malyshev~\cite{Malyshev}.

For the {\it strain gradient-like  choice}~(\ref{grad}),
the characteristic lengths~(\ref{L1}) modify to
\begin{align}
\label{L1-grad}
\ell_1^2=\ell_2^2=\frac{\alpha_1}{2\mu_e}\,,\qquad
\ell_3^2\quad \text{not defined}\,,\qquad
\ell_4^2\quad \text{not defined}\,.
\end{align}
Here, $\ell_1^2=\ell_2^2$ reproduces the characteristic length
of gradient elasticity theory of Helmholtz type (see~\cite{LM}).
Since $\mu_c\rightarrow0, \ \mu_c\geq0 $ (symmetric force stresses),
$\ell_3$ and $\ell_4$ are not defined.

Therefore, in contrast with the   Teisseyre's model \cite{Teisseyre73} (see the Subsection \ref{EChoice}),
the dislocation model proposed  by Lazar and Anastassiadis \cite{Lazar2009,Lazar2009b} is  more general
since it allows asymmetric stresses.
As we have seen, Edelen \cite{Edelen83,EdelenLagoudas,Edelen96}, Malyshev \cite{Malyshev}
and Lazar \cite{LazarJPMG02,Lazar02} discussed and used some conditions upon
the constitutive coefficients in order to obtain a model for dislocations
with symmetric force stress.
Malyshev \cite{Malyshev} and Lazar \cite{LazarJPMG02,Lazar02}
used the so-called {\it Einstein choice in three dimensions}.
One important difference between the asymmetric and the symmetric gauge theoretical models
of dislocations is that the asymmetric model possesses four characteristic
length scale parameters while a symmetric model (e.g. Einstein choice)
has only two characteristic length scale parameters~\cite{Lazar2009}.

\subsection{On the Einstein choice}\label{EChoice}
In order to write the  equations \eqref{L} in terms of the divergence operator (see Eqs. (1)-(4) and (36)-(38)  from \cite{Teisseyre73} and  Eq. (3.39) from \cite{EringenClaus}) the following {\it moment stress tensor} has been introduced
\begin{align}\label{cecet}
\Lambda_{plk}&=a_1\,\alpha_{rn}(\epsilon_{prn}\delta_{kl}-
\epsilon_{krn}\delta_{pl})+
a_2\, \epsilon_{pkn}\alpha_{ln}
+a_3\,(\epsilon_{pln}\alpha_{kn}-\epsilon_{kln}\alpha_{pn}).
\end{align}

In a previous paper \cite{NeffGhibaMicroModel} we have  identified the constitutive coefficients of the dislocation energy in the Eringen-Claus model \cite{Eringen_Claus69,EringenClaus,Eringen_Claus71}  with the coefficients in our isotropic case, namely
\begin{align}\label{omE}
\alpha_1=a_2-a_3,\quad\quad \ \alpha_2=a_2-a_3-2a_1,\quad\quad\ \ \alpha_3=\frac{2a_3+a_2}{3}.
\end{align}

Imposing the additional assumption that the moments of rotations have to vanish, Teisseyre  \cite{Teisseyre74} also requires  that  the corresponding differences between the stress moment tensor components and body couples appearing in the equation vanish. This is the reason why  Teisseyre \cite{Teisseyre74} assumed that
\begin{align}\label{lambdaT}
\Lambda_{plk,p}=\Lambda_{pkl,p}, \quad\quad \quad M_{lk}=M_{kl}\, .
\end{align}
In order to satisfy \eqref{lambdaT}$_1$,   Teisseyre considered the following sufficient condition\footnote{In fact the condition $a_2=a_1+a_3$ is necessary and sufficient to satisfy \eqref{lambdaT}$_1$ if $P\in\Sym(3)$. In addition,  in another paper \cite{Teisseyre74},  Teisseyre assumed that $a_3=0$ which removes the effects of the {\it micro-dislocation} tensor $\alpha=-\Curl P$ completely.
}
\begin{align}\label{Teiscond}
a_2=-a_3,\quad \quad\quad a_1=-2a_3\,.
\end{align}
In terms of our notations and in the dislocation gauge theory such a condition reads~\cite{LazarJPMG02,Lazar2009,LH}
\begin{align}
\label{Lcond}
\alpha_2=-\alpha_1,\quad \quad\quad \alpha_3=-\frac{\alpha_1}{6}\, ,
\end{align}
and this is called the {\it Einstein choice} in three dimensions~\cite{LazarJPMG02,LH}.

Moreover, using the assumption \eqref{lambdaT}$_1$,  it is natural to introduce the tensor
\begin{align}
m_{kl}=\frac{1}{2}\epsilon_{kmn}\,\Lambda_{mln},
\end{align}
and further, written in terms of the operator $\Curl$, it  turns into
\begin{align}\label{ma1a2}
m=&a_3\,\tr(\Curl P){\cdp} \id+2a_1\,\skew \Curl P+(a_2-a_3)\,(\Curl P)^T.
\end{align}
In terms of $m$, the equation \eqref{L}$_2$ may be rewritten as
\begin{align}\label{formatcurl}
 \widehat{\sigma}^0&=-\Curl(m^T)+2{\mu}_e  \sym(\nabla u-{P})+2{\mu_c} \skew(\nabla u-{P})+{\lambda}_e\, \tr(\nabla u-{P}){\cdp} \id\, .
\end{align}

In other words (see \cite{NeffGhibaMicroModel}), equation \eqref{lambdaT}$_1$ demands that $m$ is such that
\begin{align}\label{tconstrains}
\Curl (m^T)\in \Sym(3).
\end{align}
 Hence, in view of \eqref{ma1a2}, the  previous constraint \eqref{tconstrains} means that
\begin{align}\label{ma1a3}
\Curl[\alpha_1\dev\sym\Curl P+\alpha_2\skew \Curl P+\alpha_3 \, \tr(\Curl P) {\cdp} \id]\in \Sym(3)\, ,
\end{align}

The above symmetry condition was studied in \cite{NeffGhibaMicroModel} and the following result has been established:
\begin{remark}
\begin{itemize}\item[]
\item[i)] If  \ \ $\alpha_1=-6\,\alpha_3$ and \ \ $ \alpha_2=6\,\alpha_3$, then
    \begin{align}
\Curl \{\alpha_1 \dev \sym \Curl P+ \alpha_2 \skew \Curl P+ {\alpha_3}\ {\rm tr}(\Curl P)\!\cdot\!\id\} \in \Sym(3)\quad  \quad \forall\,  P\in\mathbb{R}^{3\times3}.
\end{align}
\item[ii)] Given $P\in\Sym(3)$, then we have
\begin{align}
\Curl \{\alpha_1 \dev \sym \Curl P+ \alpha_2 \skew \Curl P+ {\alpha_3}\ {\rm tr}(\Curl P)\!\cdot\!\id\} \in \Sym(3)
\end{align}
 if and only if $\ \alpha_1=-\alpha_2$.
\end{itemize}
\end{remark}

Thus,   the Einstein choice \eqref{Lcond} implies that
\begin{align}\label{ma1a6}
\Curl[m^T]\in \Sym(3)\,  \quad\quad \text{for all}\quad \quad P\in\mathbb{R}^{3\times3}.
\end{align}

It is obvious that the Einstein choice~(\ref{Lcond})
violates the conditions~(\ref{condpara-gt}) and (\ref{condpara}) (see also \cite{Lazar2009}).
The conditions~\eqref{Lcond} were used  by  Malyshev \cite{Malyshev} and Lazar \cite{LazarJPMG02,Lazar02}
in order to investigate dislocations with symmetric force stress\footnote{In the Lazar's
original notations the conditions \eqref{Lcond} becomes  $a_2 = -a_1$ and $a_3 = -\frac{a_1}{2}$}.
Using the Einstein choice~(\ref{Lcond}), the constitutive relation~(\ref{CR-iso})$_2$ reduce to~\cite{LazarJPMG02,Lazar02}
\begin{align}
\label{CR-EC}
m&=\alpha_1\, \kappa\,,
\end{align}
where
\begin{align}
\label{Nye}
\kappa=\alpha^T-\frac{1}{2}\,\tr(\alpha) {\cdp}\id
\end{align}
is the well-known  {\it Nye tensor} (e.g. \cite{Nye,Kroener58,Kroener81,Neff_curl06}).
The inverse is given by
\begin{align}
\label{Nye2}
\alpha=\kappa^T-\tr(\kappa) {\cdp}\id\,.
\end{align}
Then the balance of dislocation stresses ~(\ref{fe-gt})$_2$ reads
\begin{align}
\label{EC-fe}
\widehat{\sigma}^0=\widehat{\sigma}-\alpha_1\, \text{inc}\, (\sym  {e})\,,
\end{align}
where $$\text{inc}(\cdot)={\rm Curl}(({\rm Curl}\,\,\cdot\,)^T)$$ denotes the incompatibility operation, which is defined as
the ${\rm Curl}$ from the right and the ${\rm Curl}$ from the left acting on a tensor of rank two~\cite{Kroener58,Kroener81}.
The tensor $\text{inc}\, (\sym  {e})$ is equivalent to the (linearized)
three-dimensional Einstein tensor~\cite{Lazar02}.
Eq.~(\ref{EC-fe}) may be decomposed into the symmetric and the skew-symmetric parts
\begin{align}
\label{EC-fe2}
\sym \widehat{\sigma}^0&=\widehat{\sigma}-\alpha_1 \text{inc}\, (\sym  {e})\\
\skew \widehat{\sigma}^0&=\skew \widehat{\sigma}\,.\notag
\end{align}

Hence, if $\mu_c=0$, then  the Einstein choice \eqref{Lcond} implies that  $\widehat{\sigma}^0\in \Sym(3)$. Using the Einstein choice and $\mu_c=0$,
the gauge theoretical dislocation model possesses only symmetric
force stresses and no moment stresses and is described by Eq.~(\ref{EC-fe2}).

\begin{remark}
There are no general existence result for the minimization problem corresponding to the Einstein choice, since in this case the internal density energy is not positive definite.
\end{remark}

\section*{Acknowledgements}
Ionel-Dumitrel Ghiba acknowledges support from the Romanian National Authority for Scientific Research (CNCS-UEFISCDI), Project No. PN-II-ID-PCE-2011-3-0521. Markus Lazar gratefully acknowledges the grants from the
Deutsche Forschungsgemeinschaft (Grant Nos. La1974/2-2, La1974/3-1).

\bibliographystyle{plain} 
\addcontentsline{toc}{section}{References}

\begin{footnotesize}

\end{footnotesize}

\end{document}